\documentclass[fleqn,usenatbib,useAMS]{Papers}
\usepackage{newtxtext,newtxmath}
\usepackage[T1]{fontenc}
\usepackage{subfloat}
\usepackage{makecell}
\usepackage[authoryear]{natbib}
\usepackage{graphicx}	
\usepackage{amsmath}	

\title[Energy, momentum, and spin parameter in dark matter flow]{Evolution of energy, momentum, and spin parameter in dark matter flow and integral constants of motion}

\author[Z. Xu]{
Zhijie (Jay) Xu,$^{1}$\thanks{E-mail: zhijie.xu@pnnl.gov; zhijiexu@hotmail.com}
\\
$^{1}$Physical and Computational Sciences Directorate, Pacific Northwest National Laboratory; Richland, WA 99352, USA\\
}

\date{Accepted XXX. Received YYY; in original form ZZZ}

\pubyear{2022}

\begin{document}
\label{firstpage}
\pagerange{\pageref{firstpage}--\pageref{lastpage}}
\maketitle

\begin{abstract}
N-body equations of motion in comoving system and expanding background are reformulated in a transformed system with static background and fixed damping. The energy and momentum evolution in dark matter flow are rigorously formulated for both systems. The energy evolution in transformed system has a simple form that is identical to the damped harmonic oscillator. The cosmic energy equation can be easily derived in both systems. For entire N-body system, 1) combined with the two-body collapse model (TBCM), kinetic and potential energy increase linearly with time $t$ such that $K_p=\varepsilon_ut$ and $P_y=-7\varepsilon_ut/5$, where $\varepsilon_u$ is a constant rate of energy cascade; 2) an effective gravitational potential exponent $n_e=-10/7\ne-1$ ($n_e=-1.38$ from simulation) can be identified due to surface energy of fast growing halos; 3) the radial momentum $G\propto a^{3/2}$ and angular momentum $H\propto a^{5/2}$, where $a$ is the scale factor. On halo scale, 1) halo kinetic and potential energy can be modelled by two dimensionless constants $\alpha_s^*$ and $\beta_s^*$. Both constants are independent of time and halo mass; 2) both halo radial and angular momentum $\propto a^{3/2}$ and can be modeled by two mass-dependent coefficients $\tau_s^*$ and $\eta_s^*$; 3) halo spin parameter is determined by $\alpha_s^*$ and $\eta_s^*$ and decreases with halo mass with derived values of 0.09 and 0.031 for small and large halos. Finally, the radial and angular momentum are closely related to the integral constants of motion $I_m$, i.e. the integral of velocity correlation or the $m$th derivative of energy spectrum at long wavelength limit. On large scale, angular momentum is negligible, $I_2$=0 reflects the conservation of linear momentum, while $I_4$ reflects the fluctuation of radial momentum $G$. On halo scale, $I_4$ is determined by both momentum that are comparable with each other.
\end{abstract}

\begin{keywords}
\vspace*{-10pt}
Dark matter; N-body simulations; Theoretical models
\end{keywords}

\begingroup
\let\clearpage\relax
\tableofcontents
\endgroup

\section{Introduction}
\label{sec:1}
The large-scale gravitational collapse of collisionless dark matter forms the basis of standard models for structure formation. The localized, over-dense halos are fundamental structures in collisionless system \citep{Neyman:1952-A-Theory-of-the-Spatial-Distri}, both from observations and large scale simulations \citep{Evrard:2002-Galaxy-clusters-in-Hubble-volu}. Recent work also demonstrates that halo structure is necessary to form in long-range interaction systems to maximize entropy \citep{Xu:2021-The-maximum-entropy-distributi,Xu:2021-Mass-functions-of-dark-matter-}. 

The halo-mediated inverse mass/energy cascade is a key feature of self-gravitating collisionless dark matter flow (SG-CFD) \citep{Xu:2021-Inverse-mass-cascade-mass-function}: "Little halos have big halos, That feed on their mass; And big halos have greater halos, And so on to growth". Halos pass their mass onto larger and larger halos, until mass growth becomes dominant over mass propagation. This conceptual picture resembles the eddy-mediated energy cascade in hydrodynamic turbulence, i.e. eddies pass their kinetic energy to smaller and smaller eddies, until viscous effects becomes dominant over inertia effect. The energy cascade in SG-CFD is facilitated by the growth of halos and highly correlated with inverse mass cascade \citep{Xu:2021-Inverse-mass-cascade-mass-function,Xu:2021-Inverse-and-direct-cascade-of-}. By contrast, "vortex stretching" (deformation of vortex structure) enables the energy cascade in hydrodynamic turbulence \citep{Taylor:1932-The-transport-of-vorticity-and,Taylor:1938-Production-and-dissipation-of-,Xu:2021-Inverse-and-direct-cascade-of-}. 

This paper focus on the large-scale evolution of energy and momentum in SG-CFD. A brief review on the large-scale dynamics of hydrodynamic turbulence should be beneficial. While Kolmogorov's theory of the universal equilibrium range is a huge success in theory of turbulence \citep{Kolmogoroff:1941-The-local-structure-of-turbule,Kolmogoroff:1941-Dissipation-of-energy-in-the-l}, it primarily focus on the small scales. Very often, the dynamics on the large scale is of practical importance for many practical applications that involve large scale momentum and mass transfer. When large scale dynamics is concerned, turbulence can be classified into two categories: 
\begin{enumerate}
\item \noindent The first category is a forced stationary turbulence that is often studied with an external forcing term to drive and maintain the turbulence. Forced stationary turbulence can achieve higher Reynolds number and longer statistics. Kinetic energy is continuously injected at the integral length scale \textit{l} (energy injection scale), cascaded down to the smallest scale, and destroyed by viscous force. For forced stationary turbulence, the energy dissipation rate $\varepsilonup$ is a constant and total energy of turbulence is also conserved. 
\item \noindent The second category is a freely decaying turbulence that is free from any external force to maintain the turbulence. There is no energy injection on large scale and total energy is continuously decaying with time. Both integral scale \textit{l} and the energy dissipation rate $\varepsilonup$ vary with time, i.e. the dynamics on the large scale. 
\end{enumerate}

It is challenging to predict the large-scale dynamics of freely decaying turbulence. The starting point is the energy equation 
\begin{equation} 
\label{ZEqnNum277600} 
\varepsilon =\frac{du^{2} }{dt} =-A\frac{u^{3} }{l} ,           
\end{equation} 
where $u^{2}$ is the kinetic energy of entire system and $A$ is a numerical constant. Equation  \eqref{ZEqnNum277600} is not closed with two unknows $u^{2} $ and integral length scale \textit{l}. The first remarkable result is to combine the relation $u^{2} l^{5} =const$ on large scale from Loitsyansky integral constant \citep{Loitsyansky:1939-Some-basic-laws-for-isotropic-} with energy equation \eqref{ZEqnNum277600} to obtain the variation of kinetic energy $u^{2} \sim t^{{-10/7} } $, the integral scale $l\sim t^{{2/7} } $, and the energy dissipation rate $\varepsilon \sim t^{{-17/7} }$, i.e. the Kolmogorov decay laws. 

When self-gravitating collisionless system in expanding background is concerned, the energy evolution can be precisely described by a cosmic energy equation (Irvine \citep{Irvine:1961-Local-Irregularities-in-a-Univ} and Layzer \citep{Layzer:1963-A-Preface-to-Cosmogony--I--The}). Collisionless particles interacting via a Newtonian gravitational potential satisfy an energy equation,
\begin{equation} 
\label{ZEqnNum325802} 
\frac{\partial E_{y} }{\partial t} +H\left(2K_{p} +P_{y} \right)=0,         
\end{equation} 
which is a manifestation of energy conservation in expanding background (mimics Eq. \eqref{ZEqnNum277600}). Here $K_{p} $ is the peculiar kinetic energy, $P_{y} $ is the potential energy in physical coordinate that can be related to the two-point density correlation \citep{Mo:1997-Analytical-approximations-to-t, Xu:2022-The-statistical-theory-of-2nd}, $E_{y} =K_{p} +P_{y} $ is the total energy of the system, $H={\dot{a}/a} $ is the Hubble parameter, and \textit{a} is the scale factor. 

Since its discovery, the cosmic energy equation was applied for many applications include estimating matter density \citep{Peebles:1980-The-Large-Scale-Structure-of-t} and gravitational binding energy \citep{Fukugita:2004-The-cosmic-energy-inventory}. Due to the continuous formation and virilization of halos, the kinetic energy $K_{p}$ increases with time. In this regard, SG-CFD is a freely growing turbulence. Just like Eq. \eqref{ZEqnNum277600}, the cosmic energy equation \eqref{ZEqnNum325802} only provides a single relation between $K_{p} $ and $P_{y} $. Additional information is required to close the energy equation \eqref{ZEqnNum325802} for dynamics of energy evolution on large scale. 

The two-body collapse model (TBCM) can be leveraged for extra insights \citep{Xu:2021-A-non-radial-two-body-collapse}. In standard models, large-scales structures are formed from hierarchical merging of small substructures. In an infinitesimal interval $dt$, that hierarchical merging process should involve two and only two substructures. In this regard, the two-body collapse model (TBCM) in expanding background \citep{Xu:2021-A-non-radial-two-body-collapse} can be a powerful analytical tool to study the non-linear structure evolution. In this paper, the exponential energy evolution for two-body collapse \citep[see][Eqs. (92) and (93)]{Xu:2021-A-non-radial-two-body-collapse} in a transformed system (suggested by the TBCM model) will be used  to close the energy Eq. \eqref{ZEqnNum325802} for energy evolution on large scale. Similar integral constants of motion for SG-CFD are also developed for dynamics of energy and momentum on large scale.

The rest of paper is organized as follows: Section \ref{sec:2} introduces the simulation data used for this work, followed by equations of motion of \textit{N}-body system in Section \ref{sec:3}. An important equivalence is established between the original comoving system in expanding background and a transformed system in static background. The energy and momentum evolution are then formulated in both systems. Comparison with \textit{N}-body simulations for energy and momentum on both large scale and halo scale is discussed in Section \ref{sec:4}. The integral constants of motion that are relevant to the dynamics of SG-CFD on large and small scales are discussion in Section \ref{sec:5}.

\section{N-body simulations and numerical data}
\label{sec:2}
The simulation data used in this work is public available and generated from large-scale \textit{N}-body simulations carried out by the Virgo consortium. A comprehensive description of the data can be found in \citep{Frenk:2000-Public-Release-of-N-body-simul,Jenkins:1998-Evolution-of-structure-in-cold}. Current study is carried out using the simulation runs with $\Omega _{0} =1$ and the standard CDM power spectrum (SCDM) to focus on the matter-dominant (Einstein--de Sitter) gravitational collapse. The same set of data has been widely used in a number of studies from clustering statistics \citep{Jenkins:1998-Evolution-of-structure-in-cold} to the formation of cluster halos in large scale environments \citep{Colberg:1999-Linking-cluster-formation-to-l} , and testing of models for halo abundances and mass functions \citep{Sheth:2001-Ellipsoidal-collapse-and-an-im}. More details on simulation parameters are provided in Table \ref{tab:1}.

Two relevant datasets from this N-boby simulation, i.e. halo-based and correlation-based statistics of dark matter flow, can be found at Zenodo.org  \citep{Xu:2022-Dark_matter-flow-dataset-part1, Xu:2022-Dark_matter-flow-dataset-part2}, along with the accompanying presentation slides, "A comparative study of dark matter flow \& hydrodynamic turbulence and its applications" \citep{Xu:2022-Dark_matter-flow-and-hydrodynamic-turbulence-presentation}. All data files are also available on GitHub \citep{Xu:Dark_matter_flow_dataset_2022_all_files}.

\begin{table}
\caption{Numerical parameters of N-body simulation}
\begin{tabular}{p{0.25in}p{0.05in}p{0.05in}p{0.05in}p{0.05in}p{0.05in}p{0.4in}p{0.1in}p{0.4in}p{0.4in}} 
\hline 
Run & $\Omega_{0}$ & $\Lambda$ & $h$ & $\Gamma$ & $\sigma _{8}$ & \makecell{L\\(Mpc/h)} & $N$ & \makecell{$m_{p}$\\$M_{\odot}/h$} & \makecell{$l_{soft}$\\(Kpc/h)} \\ 
\hline 
SCDM1 & 1.0 & 0.0 & 0.5 & 0.5 & 0.51 & \centering 239.5 & $256^{3}$ & 2.27$\times 10^{11}$ & \makecell{\centering 36} \\ 
\hline 
\end{tabular}
\label{tab:1}
\end{table}

\section{Equations of motion and evolution of system energy and momentum}
\label{sec:3}
In this section, we focus on the evolution of energy and momentum of a \textit{N}-body system in expanding background with a power-law gravitational potential and an arbitrary potential exponent \textit{n}.\textit{ }An equivalence was established between comoving system in expanding background and a transformed system in static background \citep{Xu:2021-A-non-radial-two-body-collapse}. We will briefly review this equivalence and focus on the temporal evolution of energy and momentum of entire system.
 
\subsection{Equations of motion for comoving and transformed systems}
\label{sec:3.1}
In a comoving system with comoving coordinates $\boldsymbol{\mathrm{x}}$ and physical time \textit{t}, equations of motion for \textit{N} self-gravitating collisionless particles in expanding background read \citep{Peebles:1980-The-Large-Scale-Structure-of-t}
\begin{equation} 
\label{eq:3} 
\frac{d^{2} \boldsymbol{\mathrm{x}}_{i} }{dt^{2} } +2H\frac{d\boldsymbol{\mathrm{x}}_{i} }{dt} =-\frac{Gm_{p} }{a^{3} } \sum _{j\ne i}^{N}\frac{\boldsymbol{\mathrm{x}}_{i} -\boldsymbol{\mathrm{x}}_{j} }{\left|\boldsymbol{\mathrm{x}}_{i} -\boldsymbol{\mathrm{x}}_{j} \right|^{3} }  ,        
\end{equation} 
where \textit{N} particles have equal mass $m_{p} $ and $G$ is the standard gravitational constant. The Hubble constant $H\left(t\right)={\dot{a}/a} $, where \textit{a} is the scale factor. Without loss of generality, we assume a power-law pair potential $V_{p} $ with an arbitrary exponent of \textit{n}, i.e. $V_{p} \left(r\right){=-G_{n} m_{p}r^{n} } $. Here $G_{n} $ is a generalized gravitational constant (reduces to $G$ when $n=-1$). The generalized equations of motion for any \textit{n} read
\begin{equation} 
\label{ZEqnNum157264} 
\frac{d^{2} \boldsymbol{\mathrm{x}}_{i} }{dt^{2} } +2H\frac{d\boldsymbol{\mathrm{x}}_{i} }{dt} =\frac{nG_{n} m_{p} }{a^{3} } \sum _{j\ne i}^{N}\frac{\boldsymbol{\mathrm{x}}_{i} -\boldsymbol{\mathrm{x}}_{j} }{\left|\boldsymbol{\mathrm{x}}_{i} -\boldsymbol{\mathrm{x}}_{j} \right|^{2-n} }  .        
\end{equation} 
For matter dominant Einstein-de Sitter (EdS) model, following relations are assumed to be valid,  
\begin{equation}
H_{0}^{2} =H^{2} a^{3}, \quad \dot{a}=\frac{dH}{dt} =-\frac{3}{2} H^{2}, \textrm{and} \quad H^{2} =\frac{8\pi}{3}G\bar{\rho }_{y} \left(a\right),
\label{ZEqnNum414010}
\end{equation}

\noindent where $H_{0} $ is the Hubble constant when \textit{a}=1 and $\bar{\rho }_{y} \left(a\right)$ is the (physical) background density. In our previous work, an effective potential exponent $n_{e} \approx -1.3\ne -1$ for virial theorem has been suggested when applied to individual halos \citep[see][Eq. (96)]{Xu:2021-Inverse-mass-cascade-halo-density}. The deviation from the standard exponent of $n=-1$ comes from the continuous mass cascade and halo surface energy \citep{Xu:2021-Inverse-mass-cascade-halo-density}. In this regard, current model with an arbitrary exponent $n$ can be instructive. 

Let's introduce a transformed time scale \textit{s} as ${ds/dt} =a^{p} $, where \textit{p} is an arbitrary exponent. The original equations of motion \eqref{ZEqnNum157264} can be equivalently transformed to
\begin{equation} 
\label{ZEqnNum168751} 
\frac{d^{2} \boldsymbol{\mathrm{x}}_{i} }{ds^{2} } +\frac{d\boldsymbol{\mathrm{x}}_{i} }{ds} \left(p+2\right)a^{-p} H=\frac{nG_{n} m_{p} }{a^{3+2p} } \sum _{j\ne i}^{N}\frac{\boldsymbol{\mathrm{x}}_{i} -\boldsymbol{\mathrm{x}}_{j} }{\left|\boldsymbol{\mathrm{x}}_{i} -\boldsymbol{\mathrm{x}}_{j} \right|^{2-n} }  .      
\end{equation} 
Specifically, $p=-2$ eliminates the first order derivative and \textit{s} is the time variable for integration in \textit{N}-body simulations. By setting $p=-1$, \textit{s} is the conformal time. 

The most relevant case for this work is $p=-{3/2} $ along with the matter dominant model in Eq. \eqref{ZEqnNum414010}. For this special case, equations of motion in transformed system read (from Eq. \eqref{ZEqnNum168751})
\begin{equation} 
\label{ZEqnNum730753} 
\frac{d^{2} \boldsymbol{\mathrm{x}}_{i} }{ds^{2} } +\frac{1}{2} H_{0} \frac{d\boldsymbol{\mathrm{x}}_{i} }{ds} =nG_{n} m_{p} \sum _{j\ne i}^{N}\frac{\boldsymbol{\mathrm{x}}_{i} -\boldsymbol{\mathrm{x}}_{j} }{\left|\boldsymbol{\mathrm{x}}_{i} -\boldsymbol{\mathrm{x}}_{j} \right|^{2-n} }  =\frac{\boldsymbol{\mathrm{F}}_{i} }{m_{p} } ,      
\end{equation} 
where $\boldsymbol{\mathrm{F}}_{i} $ is the total force on particle \textit{i} from all other particles. The scale factor \textit{a} does not explicitly appear in Eq. \eqref{ZEqnNum730753} and Hubble constant $H_{0} $ can be treated as a constant damping. The original equation of motion \eqref{ZEqnNum157264} in expanding background is equivalent to Eq. \eqref{ZEqnNum730753} evolving with a new time scale \textit{s} in a static background, i.e. a transformed system with constant damping ${H_{0}/2}$. 

Starting from Eq. \eqref{ZEqnNum730753}, a non-radial two-body collapse model (TBCM) was proposed for gravitational collapse of dark matter in expanding background. For convenience, TBCM model can be analytically solved in transformed system to give the critical density of $\Delta _{c} ={2/\beta _{s2}^{2} =} 18\pi ^{2} $, where the critical value $\beta _{s2} ={1/\left(3\pi \right)} $ for equilibrium collapse \citep[see][Eq. (89)]{Xu:2021-A-non-radial-two-body-collapse}. In this regard, the TBCM model plays the same role as the harmonic oscillator in dynamics and can be fundamental to understand structure formation and evolution. 

The transformed system consists of comoving spatial coordinates $\boldsymbol{\mathrm{x}}_{i} $ and a transformed time scale \textit{s}. The particle velocity $\boldsymbol{\mathrm{v}}_{i} $ in transformed system is
\begin{equation} 
\label{ZEqnNum631619} 
\boldsymbol{\mathrm{v}}_{i} =\frac{d\boldsymbol{\mathrm{x}}_{i} }{ds} =a^{{3/2} } \frac{d\boldsymbol{\mathrm{x}}_{i} }{dt} ,           
\end{equation} 
while the peculiar velocity $\boldsymbol{\mathrm{u}}_{i} $ in physical time \textit{t} can be related to the transformed velocity $\boldsymbol{\mathrm{v}}_{i} $ as
\begin{equation} 
\label{ZEqnNum815118} 
\boldsymbol{\mathrm{u}}_{i} =a\frac{d\boldsymbol{\mathrm{x}}_{i} }{dt} =\frac{d\boldsymbol{\mathrm{r}}_{i} }{dt} -H\boldsymbol{\mathrm{r}}_{i} =a^{-{1/2} } \boldsymbol{\mathrm{v}}_{i} ,         
\end{equation} 
where $\boldsymbol{\mathrm{r}}_{i} =a\boldsymbol{\mathrm{x}}_{i} $ is the physical coordinate of particle \textit{i}.

The specific kinetic energy of transformed system (from Eq. \eqref{ZEqnNum631619}) can be related to the specific peculiar kinetic energy $K_{p} $ as 
\begin{equation} 
\label{ZEqnNum866905} 
K_{s} =\frac{1}{2N} \sum _{i=1}^{N}\boldsymbol{\mathrm{v}}_{i}^{2}  =\frac{a}{2N} \sum _{i=1}^{N}\boldsymbol{\mathrm{u}}_{i}^{2}  =aK_{p} .         
\end{equation} 

The specific potential energy $P_{s} $ of transformed system can be related to the potential $P_{y} $ in physical coordinate, where $P_{s} =a^{-n} P_{y} $ (due to $\boldsymbol{\mathrm{r}}_{i} =a\boldsymbol{\mathrm{x}}_{i} $) with
\begin{equation}
P_{s} =\frac{1}{N} \sum _{i}^{N}\phi \left(\boldsymbol{\mathrm{x}}_{i} \right) =a^{-n} P_{y} \quad \textrm{and} \quad P_{y} =\frac{1}{N} \sum _{i}^{N}\phi \left(\boldsymbol{\mathrm{r}}_{i} \right).    
\label{ZEqnNum374407}
\end{equation}
The potential $\phi \left(\boldsymbol{\mathrm{x}}_{i} \right)$ of particle \textit{i} reads
\begin{equation} 
\label{eq:12} 
\phi \left(\boldsymbol{\mathrm{x}}_{i} \right)=-\frac{1}{2} G_{n} m_{p} \sum _{j\ne i}^{N}\frac{1}{\left|\boldsymbol{\mathrm{x}}_{i} -\boldsymbol{\mathrm{x}}_{j} \right|^{-n} }  .        
\end{equation} 

The total force on particle \textit{i} is related to the gradient of potential as ${\boldsymbol{\mathrm{F}}_{i}/m_{p} } =-N{\partial P_{s} /\partial \boldsymbol{\mathrm{x}}_{i} } $. Now Eq. \eqref{ZEqnNum730753} can be rewritten as, 
\begin{equation} 
\label{ZEqnNum178889} 
\frac{d^{2} \boldsymbol{\mathrm{x}}_{i} }{ds^{2} } +\frac{1}{2} H_{0} \frac{d\boldsymbol{\mathrm{x}}_{i} }{ds} +N\frac{\partial P_{s} }{\partial \boldsymbol{\mathrm{x}}_{i} } =0.         
\end{equation} 
Since the potential energy $P_{s} $ is a function of coordinates of all particles in N-body system, the total time derivative of the potential energy $P_{s} $ can be obtained from the chain rule, 
\begin{equation} 
\label{eq:14} 
\frac{dP_{s} }{ds} =\sum _{i}^{N}\frac{\partial P_{s} }{\partial \boldsymbol{\mathrm{x}}_{i} }  \frac{d\boldsymbol{\mathrm{x}}_{i} }{ds} .          
\end{equation} 

In this section, the original equations of motion (Eq. \eqref{ZEqnNum157264}) for a comoving system in expanding background is equivalently transformed to Eq. \eqref{ZEqnNum730753} for a transformed system in static background with a constant damping. While two systems are essentially equivalent, for convenience, it is easier to formulate the energy and momentum evolution in transformed system.

\subsection{Temporal evolution of system energy }
\label{sec:3.2}
The energy evolution for entire system can be obtained by multiplying ${\boldsymbol{\mathrm{v}}_{i} =d\boldsymbol{\mathrm{x}}_{i} /ds} $ to both sides of Eq. \eqref{ZEqnNum178889}  and adding equations for all particles together. An exact and simple equation (in transformed system) for specific energy (energy per unit mass) reads
\begin{equation} 
\label{ZEqnNum333845} 
\frac{\partial E_{s} }{\partial s} +H_{0} K_{s} =0,          
\end{equation} 
where the total energy $E_{s} =K_{s} +P_{s} $ in transformed system is monotonically decreasing with time \textit{s} due to the constant damping $H_{0} $. 

Equation \eqref{ZEqnNum333845} is exact and valid for arbitrary potential exponent \textit{n} (\textit{n} is not involved in Eq. \eqref{ZEqnNum333845}).\textit{ }Note that Eq. \eqref{ZEqnNum333845} is exactly same as the energy equation for a damped single-degree-of-freedom harmonic oscillator, which has been well-studied with well-known exponential solutions. With two unknowns ($E_{s} $ and $K_{s} $) in Eq. \eqref{ZEqnNum333845}, additional information from a TBCM model will be used for a complete solution of Eq. \eqref{ZEqnNum333845} (see Section \ref{sec:4.1}). 

Equation \eqref{ZEqnNum333845} can be transformed to an equation with respect to the scale factor \textit{a} using the relation ${da/ds} =H_{0} a$ (from ${ds/dt} =a^{p} $ with $p=-{3/2} $) ,
\begin{equation} 
\label{ZEqnNum599555} 
\frac{\partial \left(K_{s} +P_{s} \right)}{\partial a} +\frac{K_{s} }{a} =0.          
\end{equation} 
Substitution of the peculiar kinetic energy $K_{p} $ (Eq. \eqref{ZEqnNum866905}) and the physical potential energy $P_{y} $ (Eq. \eqref{ZEqnNum374407}) into Eq. \eqref{ZEqnNum599555} leads to the energy evolution in the original comoving system with an arbitrary potential exponent \textit{n},
\begin{equation} 
\label{ZEqnNum727316} 
\frac{\partial \left[a\left(K_{p} +a^{-n-1} P_{y} \right)\right]}{\partial a} +K_{p} =0,        
\end{equation} 
which exactly reduces to the cosmic energy Eq. \eqref{ZEqnNum325802} by setting $n=-1$. We can equivalently present the energy evolution with respect to the physical time \textit{t} as,  
\begin{equation} 
\label{ZEqnNum769508} 
\frac{\partial \left(K_{p} +a^{-n-1} P_{y} \right)}{\partial t} +H\left(2K_{p} +a^{-n-1} P_{y} \right)=0.       
\end{equation} 

A more convenient form is to normalize the kinetic and potential energy by a one-dimensional peculiar velocity dispersion $u_{}^{2} \left(a\right)={2/3} K_{p} \left(a\right)$. The original energy Eq. \eqref{ZEqnNum727316} can be rewritten as
\begin{equation} 
\label{ZEqnNum262503} 
\frac{\partial }{\partial \ln a} \ln \left(\frac{3}{2} +\frac{P_{s} }{au_{}^{2} } \right)+\frac{\partial \ln u_{}^{2} }{\partial \ln a} +1+{{3/2}\Bigg/\left(\frac{3}{2} +\frac{P_{s} }{au_{}^{2} } \right)} =0.     
\end{equation} 
For a power-law form of velocity dispersion $u_{}^{2} \left(a\right)\propto a^{\gamma } $ (solutions of TBCM model justifies a power-law evolution, see Section \ref{sec:4.1}), Eq. \eqref{ZEqnNum262503} admits an exact steady-state solution for the ratio between potential and kinetic energies,
\begin{equation} 
\label{ZEqnNum298738} 
\frac{a^{-n} P_{y} }{aK_{p} } =\frac{P_{s} }{K_{s} } =\frac{2}{3} \frac{P_{s} }{au^{2} } =-\frac{\gamma +2}{\gamma +1} .        
\end{equation} 

The total energy ($E_{y} =K_{p} +P_{y}$) and the virial energy (defined as $V_{y} =2K_{p} -nP_{y} $) normalized by the kinetic energy $K_{p} $ are,
\begin{equation}
\begin{split}
&\frac{E_{y} }{K_{p} } =\frac{K_{p} +P_{y} }{K_{p} } =1+\left(-\frac{\gamma +2}{\gamma +1} \right)a^{1+n}\\
&\textrm{and}\\  
&\frac{V_{y} }{K_{p} } =\frac{2K_{p} -nP_{y} }{K_{p} } =2-n\left(-\frac{\gamma +2}{\gamma +1} \right)a^{1+n}.
\label{eq:21}
\end{split}
\end{equation}
\noindent Specifically, for standard gravitational interaction with $n=-1$, 
\begin{equation}
\frac{E_{y} }{K_{p} } =\frac{K_{p} +P_{y} }{K_{p} } =-\frac{1}{\gamma +1},\quad \frac{V_{y} }{K_{p} } =\frac{2K_{p} -nP_{y} }{K_{p} } =\frac{\gamma }{\gamma +1},   
\label{ZEqnNum259775}
\end{equation}

\noindent where $\gamma =0$ corresponds to a static system with constant energy. 

For a dynamically evolving \textit{N}-body system, the total energy and virial energy normalized by $K_{p} \left(a\right)$ or $u_{}^{2} \left(a\right)$ are functions of $\gamma $ only at the steady state. An effective exponent $n_{e} $ can be introduced for entire system
\begin{equation}
\begin{split}
&n_{e} =\frac{2K_{p} }{P_{y} } =-\frac{2\left(1+\gamma \right)}{2+\gamma } a^{-\left(1+n\right)}\\ 
&\textrm{and}\\
&n_{e} =-\frac{2\left(1+\gamma \right)}{2+\gamma } \quad \textrm{for}\quad n=-1,   
\end{split}
\label{ZEqnNum539248}
\end{equation}
where the standard virial theorem is satisfied in the form of $2K_{p} -n_{e} P_{y} =0$. The effective exponent $n_{e} $ equals \textit{n} if and only if $n=-1$ and $\gamma =0$. However, the kinetic energy $K_{p} \left(a\right)$ is usually increasing with time for self-gravitating collisionless flow (i.e. $\gamma \ne0$ and $V_{y} \ne 0$). Therefore, for a self-gravitating \textit{N}-body system, the effective potential exponent $n_{e} $ can be significantly different from -1 ($n_{e} \ne -1$) due to the dynamic effects of mass cascade (mass accretion) and halo surface energy \citep{Xu:2021-Inverse-mass-cascade-halo-density}. 

Equation \eqref{ZEqnNum333845} relates the total energy $E_{s} $ with the kinetic energy $K_{s} $ for a transformed system. To solve for energy evolution, another relation is required to close Eq. \eqref{ZEqnNum333845} where the elementary problem of gravitational collapse (two-body collapse) provides some additional insights \citep{Xu:2021-A-non-radial-two-body-collapse}. Analytical solutions by solving a two-body collapse model (TBCM) suggest a power-law evolution of energy for self-gravitating systems \citep[see][Eq. (92) and (93)]{Xu:2021-A-non-radial-two-body-collapse}. Section \ref{sec:4.1} presents more discussion and detail comparison with N-body simulations. 
 
\subsection{Temporal evolution of virial quantity and angular momentum}
\label{sec:3.3}
The virial theorem in expanding background can be obtained by multiplying $\boldsymbol{\mathrm{x}}_{i} $ (dot product) to both sides of Eq. \eqref{ZEqnNum178889} for all particles and taking the average,
\begin{equation} 
\label{ZEqnNum639984} 
\frac{1}{N} \sum _{i}^{N}\frac{d\boldsymbol{\mathrm{v}}_{i} }{ds}  \cdot \boldsymbol{\mathrm{x}}_{i} +\frac{1}{2} H_{0} G_{s} +\sum _{i}^{N}\frac{\partial P_{s} }{\partial \boldsymbol{\mathrm{x}}_{i} } \cdot \boldsymbol{\mathrm{x}}_{i}  =0,       
\end{equation} 
where the virial quantity (the radial momentum analogue to angular momentum) of entire system is defined as 
\begin{equation} 
\label{ZEqnNum446862} 
\begin{split}
&G_{s} =\frac{1}{N} \sum _{i}^{N}\boldsymbol{\mathrm{v}}_{i} \cdot \boldsymbol{\mathrm{x}}_{i}  =a^{{1/2} } G_{p} =a^{{1/2} } \frac{1}{N} \sum _{i}^{N}\boldsymbol{\mathrm{u}}_{i} \cdot \boldsymbol{\mathrm{x}}_{i}\\ 
&\textrm{and}\\
&G_{p} =\frac{1}{N} \sum _{i}^{N}\boldsymbol{\mathrm{u}}_{i} \cdot \boldsymbol{\mathrm{x}}_{i} 
\end{split}
\end{equation} 
for transformed and comoving systems, respectively. Equation for virial quantity $G_{s}$ reads (from Eq. \eqref{ZEqnNum639984})
\begin{equation} 
\label{ZEqnNum257929} 
\frac{dG_{s} }{ds} +\frac{1}{2} H_{0} G_{s} =2K_{s} +\frac{1}{N} \sum _{i}^{N}\frac{d\boldsymbol{\mathrm{v}}_{i} }{ds}  \cdot \boldsymbol{\mathrm{x}}_{i} .        
\end{equation} 

For a power-law form of potential, we have the identity,
\begin{equation} 
\label{ZEqnNum814295} 
\begin{split}
&-N\sum _{i}^{N}\frac{\partial P_{s} }{\partial \boldsymbol{\mathrm{x}}_{i} } \cdot \boldsymbol{\mathrm{x}}_{i}  =\frac{1}{m_{p} } \sum _{i}^{N}\boldsymbol{\mathrm{F}}_{i} \cdot \boldsymbol{\mathrm{x}}_{i}\\
&\quad\quad\quad =\frac{1}{m_{p} } \sum _{i}^{N}\sum _{j\ge i}^{N}\boldsymbol{\mathrm{F}}_{ij} \cdot \left(\boldsymbol{\mathrm{x}}_{j} -\boldsymbol{\mathrm{x}}_{i} \right)  =-n\sum _{i}^{N}\phi \left(\boldsymbol{\mathrm{x}}_{i} \right),
\end{split}
\end{equation} 
where the second equality assumes that Newton's third law of motion holds, i.e., $\boldsymbol{\mathrm{F}}_{ij} =-\boldsymbol{\mathrm{F}}_{ji} $. For N-body systems with periodic boundary, extra care is needed for the second equality in Eq. \eqref{ZEqnNum814295} because of periodicity. In fact, the time evolution of virial quantity $G_{p}$ in a N-body system can be related to the integral constant of motion on large scale (see Eq. \eqref{ZEqnNum374769} and Fig. \ref{fig:7} in Section \ref{sec:5.1}). 

For systems with an open boundary (unbounded) and power-law potential, equation for virial quantity $G_{s} $ can be finally written as (from Eq. \eqref{ZEqnNum639984} and using identity \eqref{ZEqnNum814295}),
\begin{equation}
\label{ZEqnNum672411} 
\frac{dG_{s} }{ds} +\frac{1}{2} H_{0} G_{s} =2K_{s} -nP_{s} .         
\end{equation} 
This is the virial theorem for self-gravitating system in transformed system with a constant damping. Equation \eqref{ZEqnNum672411} can be easily transformed back to the original comoving system as
\begin{equation}
\label{ZEqnNum724192} 
\frac{dG_{p} }{dt} +HG_{p} =\frac{2aK_{p} -na^{-n} P_{y} }{a^{2} } .        
\end{equation} 
The virial quantity $G_{s} $ and $G_{p} $ of entire system are relevant to the translational motion of halos, as we will discuss in Section \ref{sec:4.3}. 

Similarly, we can also multiply $\boldsymbol{\mathrm{x}}_{i} \times $ (cross product) to both sides of Eq. \eqref{ZEqnNum178889} for all particles to obtain the evolution of specific angular momentum (per unit mass), 
\begin{equation} 
\label{eq:30} 
\frac{d\boldsymbol{\mathrm{H}}_{s} }{ds} +\frac{H_{0} }{2} \boldsymbol{\mathrm{H}}_{s} =-\sum _{i}^{N}\boldsymbol{\mathrm{x}}_{i} \times \frac{\partial P_{s} }{\partial \boldsymbol{\mathrm{x}}_{i} }  ,         
\end{equation} 
where, 
\begin{equation}
\begin{split}
\boldsymbol{\mathrm{H}}_{s} =a^{{1/2} } \boldsymbol{\mathrm{H}}_{p} =\frac{1}{N} \sum _{i}^{N}\boldsymbol{\mathrm{x}}_{i} \times \boldsymbol{\mathrm{v}}_{i}\quad \textrm{and}  \quad \boldsymbol{\mathrm{H}}_{p} =\frac{1}{N} \sum _{i}^{N}\boldsymbol{\mathrm{x}}_{i} \times \boldsymbol{\mathrm{u}}_{i}.
\end{split}
\label{ZEqnNum346758}
\end{equation}
Here $\boldsymbol{\mathrm{H}}_{s} $ and $\boldsymbol{\mathrm{H}}_{p} $ are the specific angular momentum in transformed and comoving systems, respectively. 

For systems with an open boundary (unbounded), 
\begin{equation} 
\label{ZEqnNum475585} 
\begin{split}
\frac{d\boldsymbol{\mathrm{H}}_{s} }{ds} +\frac{H_{0} }{2} \boldsymbol{\mathrm{H}}_{s}&=-\sum _{i}^{N}\boldsymbol{\mathrm{x}}_{i} \times \frac{\partial P_{s} }{\partial \boldsymbol{\mathrm{x}}_{i} }  =\frac{1}{m_{p} } \sum _{i}^{N}\boldsymbol{\mathrm{x}}_{i} \times \boldsymbol{\mathrm{F}}_{i}\\
&=\frac{1}{m_{p} } \sum _{i}^{N}\sum _{j\ge i}^{N}\left(\boldsymbol{\mathrm{x}}_{i} -\boldsymbol{\mathrm{x}}_{j} \right)\times \boldsymbol{\mathrm{F}}_{ij}=0,
\end{split}
\end{equation} 
where the angular momentum exponentially decays ($\boldsymbol{\mathrm{H}}_{s} \sim \exp \left(-{H_{0} s/2} \right)$) in transformed system (or $\boldsymbol{\mathrm{H}}_{p} \sim a^{{-1/2} } $ in original comoving system using ${ds/dt} =a^{-{3/2} } $ in Eq. \eqref{ZEqnNum168751}). Similarly, extra care is needed for N-body systems with periodic boundaries for the third equality in Eq. \eqref{ZEqnNum475585} because of the periodicity. The time evolution of $\boldsymbol{\mathrm{H}}_{p} $ in a N-body system is modelled in Eq. \eqref{ZEqnNum748159} and plotted in Fig. \ref{fig:7}. The evolution of virial quantity and angular momentum on halo scale is discussed in Section \ref{sec:4.3}, where we show that $\boldsymbol{\mathrm{H}}_{p} $ is related to the angular momentum of halos.

\section{Evolution of energy and momentum from N-body simulation}
\label{sec:4}
\subsection{Evolution of energy on large scale}
\label{sec:4.1}
The evolution of system energy is described by Eqs. \eqref{ZEqnNum333845} and \eqref{ZEqnNum727316} for self-gravitating collisionless system. Still, Eq. \eqref{ZEqnNum333845} is not closed and we need extra information to solve it. It is well known that energy exponentially decays for a damped harmonic oscillator of single degree of freedom. The rate of decay is proportional to the damping coefficient $\alpha _{d} $, i.e. $E\propto \exp \left(-\alpha _{d} t\right)$. The exact analogue of harmonic oscillator model in self-gravitating collisionless system is a two-body collapse model (TBCM) \citep{Xu:2021-A-non-radial-two-body-collapse}. 

Solutions obtained from TBCM model suggest that both kinetic and potential energy evolve exponentially in transformed system (see Eqs. (73), (92), and (93), (94) in \citep{Xu:2021-A-non-radial-two-body-collapse}). Just like the dynamics of large structure system can be represented by coupled harmonic oscillators, the dynamics of self-gravitating collisionless system should resemble that of a simple two-body collapse. We reasonably expect the energy of entire system on large scale should evolve in the same way as two-body collapse from TBCM model, i.e. an exponential evolution in transformed system. With this in mind, the only possible solutions for Eq. \eqref{ZEqnNum333845} are
\begin{equation}
K_{s} =\alpha \exp \left(-\frac{H_{0} s}{1+{\beta /\alpha } } \right) \quad \textrm{and} \quad P_{s} =\beta \exp \left(-\frac{H_{0} s}{1+{\beta /\alpha } } \right),    
\label{ZEqnNum412772}
\end{equation}

\noindent where $\alpha $ and $\beta $ are two numerical constants.

The temporal evolution of kinetic and potential energy in transformed time scale \textit{s} (Eq. \eqref{ZEqnNum412772}) can be equivalently translated to a power-law evolution in physical time \textit{t} with relation $s=t_{0} \ln \left({t/t_{i} } \right)$ (from ${ds/dt} =a^{-{3/2} } $ in Eq. \eqref{ZEqnNum168751}), such that  
\begin{equation} 
\label{ZEqnNum840626} 
K_{p} =\varepsilon _{u} t^{-\frac{2}{3} \frac{\left(2+{\beta /\alpha } \right)}{\left(1+{\beta /\alpha } \right)} } =\varepsilon _{u} a^{-\frac{\left(2+{\beta /\alpha } \right)}{\left(1+{\beta /\alpha } \right)} } ,         
\end{equation} 
\begin{equation} 
\label{ZEqnNum404986} 
P_{s} =\frac{\beta }{\alpha } \varepsilon _{u} t^{-\frac{2}{3\left(1+{\beta /\alpha } \right)} } =\frac{\beta }{\alpha } \varepsilon _{u} a^{-\frac{1}{\left(1+{\beta /\alpha } \right)} } ,         
\end{equation} 
and
\begin{equation} 
\label{ZEqnNum680744} 
P_{y} =a^{n} P_{s} =\frac{\beta }{\alpha } \varepsilon _{u} a^{n-\frac{1}{\left(1+{\beta /\alpha } \right)} } ,         
\end{equation} 
where the constant $\varepsilon _{u} $ is a rate of production of kinetic energy or the energy flux across scales. Obviously, exponent $\gamma $ for velocity dispersion of the entire system ($u_{}^{2} \propto a^{\gamma } $) in Eq. \eqref{ZEqnNum298738} reads 
\begin{equation}
\label{ZEqnNum181262} 
\gamma =-\frac{\left(2+{\beta/\alpha } \right)}{\left(1+{\beta/\alpha } \right)} .          
\end{equation} 
As shown in Eqs. \eqref{ZEqnNum262503}, \eqref{ZEqnNum298738}, and \eqref{ZEqnNum259775}, for a constant exponent $\gamma $, the total energy normalized by $K_p$ or velocity dispersion $u_{}^{2} $ is time-invariant. In this regard, the self-gravitating collisionless system is energy conserved (if normalized by dispersion $u_{}^{2} $) that mimics the forced stationary turbulence with a conserved total kinetic energy.

By choosing an effective potential exponent $n_{e} $ for self-gravitating system (Eq. \eqref{ZEqnNum840626} and \eqref{ZEqnNum680744}) to satisfy the virial theorem, i.e. $2K_{p} =n_{e} a^{-n-1} P_{y} $ from Eq. \eqref{ZEqnNum724192}, we must have
\begin{equation} 
\label{ZEqnNum751621} 
n_{e} =\frac{2K_{p} }{a^{-n-1} P_{y} } =2\frac{\alpha }{\beta } .           
\end{equation} 
The exponent $\gamma $ for velocity dispersion $u^{2} \propto a^{\gamma } $ can be finally related to the effective potential exponent $n_{e} $ (found from Eq. \eqref{ZEqnNum181262}), 
\begin{equation}
\label{ZEqnNum259367} 
\gamma =-\frac{2\left(1+n_{e} \right)}{\left(2+n_{e} \right)} .          
\end{equation} 

The energy evolution for self-gravitating collisionless system in expanding background with a potential exponent \textit{n} can be equivalent to the same system with an effective potential exponent for virial theorem such that $n_{e} \ne n$. Since a positive scaling exponent $\gamma >0$ is expected for system with increasing kinetic energy, the effective potential exponent $n_{e} $ should be in the range of $n_{e} \in \left(-2,-1\right)$. Obviously, $\gamma =0$ is only possible for an isolated system in static background ($H_{0} =0$ in Eq. \eqref{ZEqnNum412772}), where both kinetic and potential energies are time-invariant and conserved. 

Figure \ref{fig:1a} plots the variation of kinetic and potential energies with scale factor \textit{a} from a \textit{N}-body simulation in Section \ref{sec:2}. Both energies approach a power-law scaling with scale factor \textit{a}, i.e. $K_{p} \left(a\right)\propto a^{{3/2} } $ (dominated by halo scale dynamics), after an initial period with $K_{p} \left(a\right)\propto a$ (dominated by dynamics on large scale in the linear regime before halos are formed). The inset plot presents the time variation of effective exponent $n_{e} $ that approaches a constant value of $n_{e} \approx -1.38$ (matching the effective potential exponent for halos \citep[see][Eq. (96)]{Xu:2021-Inverse-mass-cascade-halo-density}). This corresponds to a value of $\gamma \approx {3/2}$ (using Eq. \eqref{ZEqnNum259367}) that is consistent with N-body simulation. The total energy $E_{y} $ in comoving system is not conserved due to virilization.

\begin{subfigures}
\begin{figure}
\includegraphics*[width=\columnwidth]{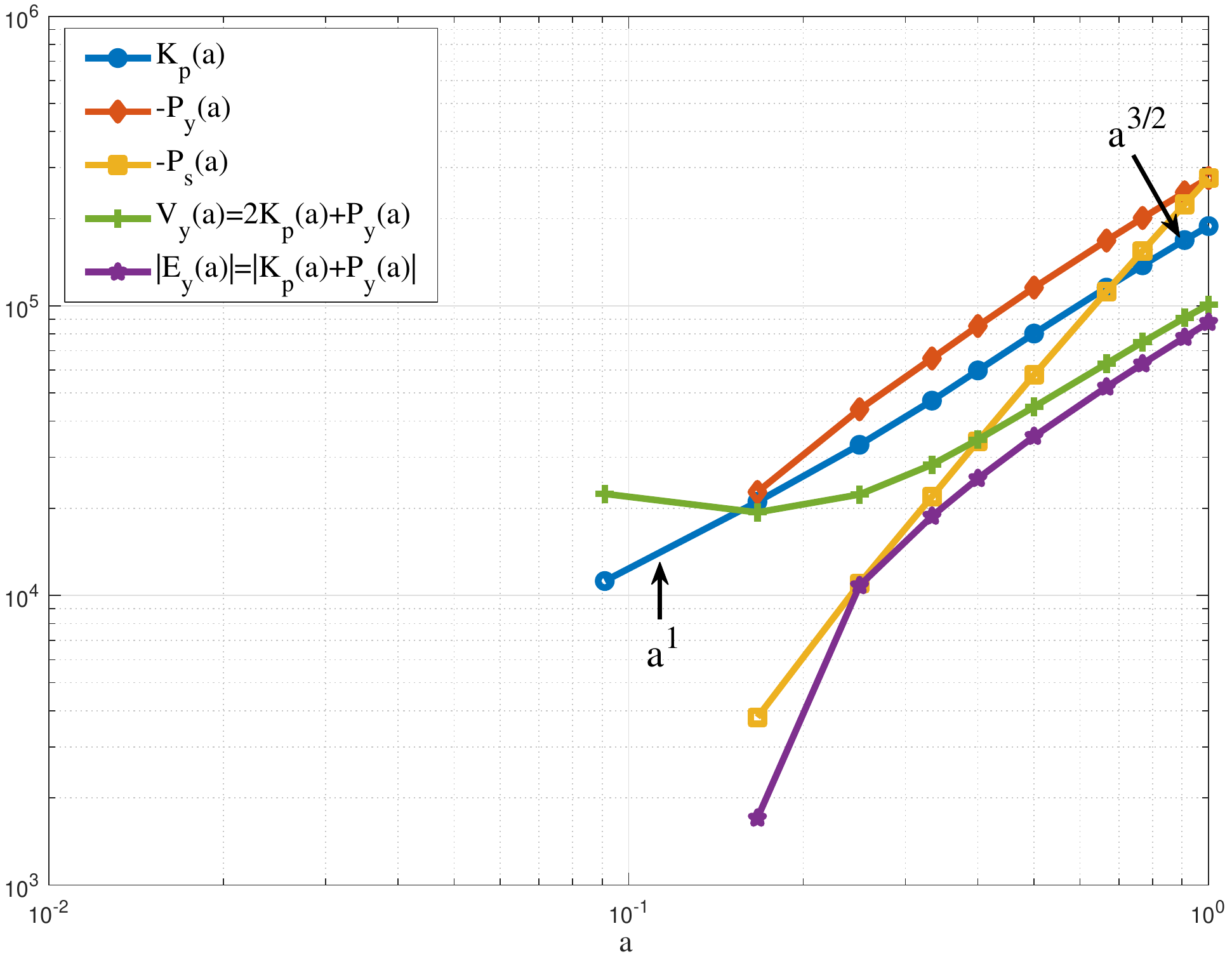}
\caption{The variation of specific kinetic and potential energies $(km^2/s^2)$ with scale factor \textit{a} from a \textit{N}-body simulation. Both energies exhibit a power-law scaling with scale factor \textit{a} at statistically steady state, with an approximate scaling of $K_{p} \left(a\right)\propto a^{{3/2} } $ and $P_{y} \left(a\right)\propto a^{{3/2} }$. The total energy $E_{y} $ in comoving system is not conserved. The virial energy $V_{y} \left(a\right)\propto a^{{3/2} } $ follows the same scaling.}
\label{fig:1a}
\end{figure}

\begin{figure}
\includegraphics*[width=\columnwidth]{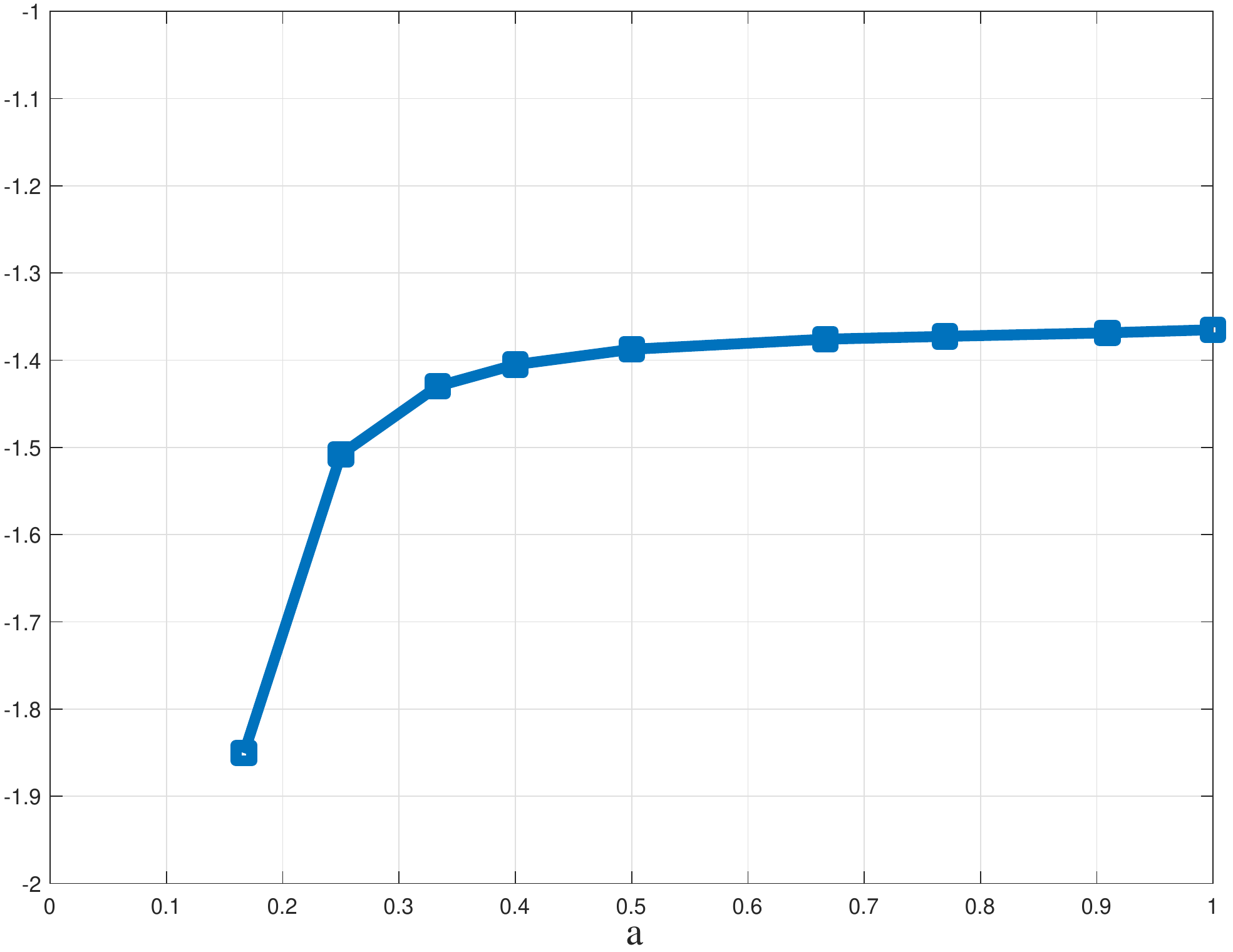}
\caption{The variation of effective exponent $n_{e} ={2K_{p}/P_{y} } $ that approaches to a constant value of -1.38, dominated by dynamics on halo scale (due to mass cascade and halo surface energy).}
\label{fig:1b}
\end{figure}
\end{subfigures}
The effective potential exponent of entire system $n_{e} \approx -1.38\ne -1$ is due to the mass cascade and halo surface energy \citep[see][Eq. (96)]{Xu:2021-Inverse-mass-cascade-halo-density}. Another interpretation is also presented here. For large halos with a typical mass $m_{h}^{L} $ that grow with a constant waiting time, the virial dispersion $\sigma _{v}^{2} (m_{h}^{L} ,a)\propto a^{-1} (m_{h}^{L} )^{{3/2} }$. For mass of that typical halo grows as $m_{h}^{L} \propto a^{{3/2} } $ \citep[see][Eq. (50)]{Xu:2021-Inverse-mass-cascade-mass-function}, the velocity dispersion $\sigma _{v}^{2} \propto a^{0} $ and the halo virial kinetic energy grows as $m_{h}^{L} \sigma _{v}^{2} \propto a^{{3/2} } $. Now consider a closed isolated halo with a fixed mass $m_h$ and the same virial kinetic energy (i.e. $m_h u^2 \propto a^{3/2}$), the velocity dispersion of that halo should scale as $u^{2} \propto a^{{3/2}}$. For such a closed isolated halo, the exponent $\gamma ={3/2}$ leads to an effective potential exponent $n_{e} =-{10/7} \approx -1.43$ from Eq. \eqref{ZEqnNum259367} that is close to $n_{e} =-1.38$ from N-body simulation. Note that $n_{e} =-{10/7} $ is one of the possible potential exponents that leads to the existence of critical value $\beta _{s2}$ in TBCM model \citep[see][Eqs. (82) and (83)]{Xu:2021-A-non-radial-two-body-collapse} and Table \ref{tab:2}). Interestingly, the energy evolution of a growing halo in N-body system with normal gravity $n=-1$ is equivalent to that of an isolate halo with an effective potential exponent $n_e=-10/7$. Table \ref{tab:2} lists several possible values of $\gamma $ along with the corresponding values of $n_{e} $ and ${\beta/\alpha } $from Eqs. \eqref{ZEqnNum751621} and \eqref{ZEqnNum259367}. Possible values of $n_{e} $ is from Equation $n_{e} ={\left(2-6m\right)/\left(1+3m\right)}$ \citep[see][Eq. (83)]{Xu:2021-A-non-radial-two-body-collapse}. 

For the special case with $\gamma ={3/2} $, we expect ${\beta/\alpha } =-{7/5} $ and $n_{e} =-{10/7} $ (from Eqs. \eqref{ZEqnNum181262} and \eqref{ZEqnNum751621}) once the statistically steady state of SG-CFD is established, where the kinetic and potential energy should evolve as
\begin{equation}
K_{s} =\varepsilon _{u}at, \quad P_{s} =-\frac{7}{5} \varepsilon _{u}at ,\quad K_{p} =\varepsilon _{u}t, \quad P_{y} =-\frac{7}{5} \varepsilon _{u}t   
\label{ZEqnNum160091}
\end{equation}
for transformed and original comoving systems, respectively. Again, $\varepsilon _{u} $ is the rate of energy production on the smallest scale for energy cascade in SG-CFD, where $\varepsilon _{u} =-{3\left(\varepsilon _{kv} +\varepsilon _{kh} \right)/2} $ with contributions from both halo virial dispersion ($\varepsilon _{kv} $) and halo velocity dispersion ($\varepsilon _{kh} $) \citep[see][Eqs. (27), (47), Figs. 3 and 4]{Xu:2021-Inverse-and-direct-cascade-of-}. 

The peculiar kinetic energy $K_{p} $ increases proportionally to the physical time \textit{t} ($K_{p} \propto a^{{3/2} } \propto t$) such that the rate of energy production $\varepsilon _{u} $ can be estimated to be,  
\begin{equation}
\label{ZEqnNum539584} 
\varepsilon _{u} =-\frac{3}{2} \frac{\partial u^{2} }{\partial t} \approx -\frac{3}{2} \frac{u_{0}^{2} }{t_{0} } =-\frac{9}{4} u_{0}^{2} H_{0} \approx -4.6\times 10^{-7} \frac{m^{2} }{s^{3} }  
\end{equation} 
for $u_{0} =354.61{km/s} $ from the simulation in Section \ref{sec:2}. The rate of energy cascade $\epsilon_u$ is a key constant can be applied to postulate dark matter particle mass and properties \citep{Xu:2022-Postulating-dark-matter-partic}, interpret the origin of critical MOND (Modified Newtonian Dynamics) acceleration $a_0$ \citep{Xu:2022-The-origin-of-MOND-acceleratio}, and derive the baryonic-to-halo mass ratio \citep{Xu:2022-The-baryonic-to-halo-mass-rela}.

\begin{table}
\begin{tabular}{p{0.2in}p{0.3in}p{0.3in}p{0.3in}p{0.3in}p{0.3in}p{0.3in}} \hline 
$\gamma $ & 0 & \textbf{3/2} & 3 & 9/2 & 6 & $\mathrm{\infty}$ \\ \hline
$n_{e} $ & -1 & \textbf{-10/7} & -8/5 & -22/13 & -7/4 & -2 \\ \hline 
${\beta/\alpha } $ & -2 & \textbf{-7/5} & -4/5& -11/13 & -7/8 & -1 \\ \hline 
\end{tabular}
\caption{List of possible parameters $\gamma$, $n_{e}$ and ${\beta/\alpha}$ for an isolate halo based on a two-body collapse model \citep[see][Eqs. (82) and (83)]{Xu:2021-A-non-radial-two-body-collapse}.}
\label{tab:2}
\end{table}

\subsection{Evolution of energy on halo scale}
\label{sec:4.2}
This section focuses on the energy evolution on the scale of halos. In N-body simulation, we first compute the (physical) root mean square radius $r_{g} $ about the center of mass of a given halo and the one-dimensional peculiar velocity dispersion $\sigma _{v}^{2} $ for every halo identified,
\begin{equation}
\begin{split}
&r_{g} =\sqrt{{\sum _{k=1}^{n_{p} }\left(m_{p} r_{k}^{2} \right)/\left(n_{p} m_{p} \right)} },\\
&\sigma _{v}^{2} =\frac{1}{3n_{p} } \sum _{k=1}^{n_{p} }\left|\boldsymbol{\mathrm{u}}_{k}^{} -\boldsymbol{\mathrm{u}}_{h}^{} \right|^{2} \quad \textrm{and} \quad \boldsymbol{\mathrm{u}}_{h} =\frac{1}{n_{p} } \sum _{k=1}^{n_{p} }\boldsymbol{\mathrm{u}}_{k},
\end{split}
\label{ZEqnNum753014}
\end{equation}

\noindent where $n_{p} $ is the number of particles in a halo, $m_{p} $ is the particle mass, $r_{k} $ is the physical distance of \textit{k}th particle to the center of mass of halo that particle resides in. Here $\boldsymbol{\mathrm{u}}_{k} $ is the particle peculiar velocity and $\boldsymbol{\mathrm{u}}_{h} $ is the halo peculiar velocity as the mean velocity of all particles in the same halo. 

Both root mean square radius $r_{g} $ and halo virial dispersion $\sigma _{v}^{2} $ can be easily computed for each halo identified in simulation. In addition, both quantities can be analytically computed for halos with a given density profile. Typical examples are a power-law density profile and a Navarro--Frenk--White (NFW) profile \citep{Navarro:1997-A-universal-density-profile-fr}:  

\begin{enumerate}
\item  \noindent For spherical halos of size $r_{h}$ with a power-law density profile of $\rho _{h} \left(r\right)\sim r^{-m} $, the root mean square radius can be found as
\begin{equation}
\label{ZEqnNum229152} 
r_{g} =\gamma _{g} r_{h} =\sqrt{\frac{\int _{0}^{r_{h} }x^{2} \rho _{h} \left(x\right)4\pi x^{2} dx }{\int _{0}^{r_{h} }\rho _{h} \left(x\right)4\pi x^{2} dx } } =\sqrt{\frac{3-m}{5-m} } r_{h} ,       
\end{equation} 
where $\gamma_{g} $ is a constant to relate $r_{g} $ to halo virial size $r_{h} $. Radius of gyration of a spherical halo is \citep[see][Eq. (75)]{Xu:2022-The-mean-flow--velocity-disper}
\begin{equation} 
\label{eq:44} 
r_{rg} =r_{h} \sqrt{\frac{2\left(3-m\right)}{3\left(5-m\right)} } =r_{g} \sqrt{\frac{2}{3} } .         
\end{equation} 

The potential energy $\Phi _{h} $ of that halo can be computed as,
\begin{equation} 
\label{ZEqnNum260973}
\begin{split}
&\Phi _{h}=-\gamma _{\Phi } \frac{Gm_{h} }{r_{h} }\\
&=-\frac{\int _{0}^{r_{h} }\frac{G}{y}  \left[\int _{0}^{y}\rho _{h} \left(x\right)4\pi x^{2} dx \right]\rho _{h} \left(y\right)4\pi y^{2} dy}{\int _{0}^{r_{h} }\rho _{h} \left(x\right)4\pi x^{2} dx } =-\frac{3-m}{5-2m} \frac{Gm_{h} }{r_{h} },
\end{split}
\end{equation} 
where $\gamma _{\Phi } $ is a constant for halo potential that is on the order of one. 

From virial theorem, $3\sigma _{v}^{2} -n_{s}^{*} \Phi _{h} =0$  ($n_{s}^{*} $ is an effective potential exponent defined for halos of different size), the one-dimensional virial velocity dispersion is (with virial ratio $\gamma _{v} =-n_{s}^{*} $), 
\begin{equation}
\label{ZEqnNum117366} 
\sigma _{v}^{2} =\frac{n_{s}^{*} }{3} \Phi _{h} =\frac{\gamma _{v} }{3} \left(\frac{3-m}{5-2m} \right)\frac{Gm_{h} }{r_{h} } ,        
\end{equation} 
where $\gamma _{v} $ is a virial ratio. Specifically, for a singular isothermal sphere profile with $m=2$, 
\begin{equation}
r_{g} =\frac{r_{h} }{\sqrt{3} } \quad \textrm{and} \quad \sigma _{v}^{2} =-\frac{\gamma _{v} }{3} \frac{Gm_{h} }{r_{h} } =\frac{\gamma _{v} }{3\sqrt{3} } \frac{Gm_{h} }{r_{g} },     
\label{eq:47}
\end{equation}

\noindent where $n_{s}^{*} =-\gamma _{v} =-1.5$ for an isothermal density profile \citep[see][Eq. (96)]{Xu:2021-Inverse-mass-cascade-halo-density} or \citep[see][Table 3]{Xu:2022-The-mean-flow--velocity-disper}). 
\\
\item \noindent Alternatively, for halos with a NFW density profile, a concentration parameter \textit{c}, size $r_{h}$, the root mean square radius is (similar to Eq. \eqref{ZEqnNum229152} and for $c=4\sim 10$), 
\begin{equation}
\label{ZEqnNum159016} 
\begin{split}
r_{g} =\gamma _{g} r_{h} &=\frac{r_{h} }{2} \sqrt{\frac{2c\left(c^{2} -3c-6\right)+12\left(1+c\right)\ln \left(1+c\right)}{c^{2} \left(1+c\right)\ln \left(1+c\right)-c^{3} } }\\
&\approx \left(0.56\sim 0.40\right)r_{h}.
\end{split}
\end{equation} 

The total potential energy $\Phi _{h} $ of a NFW halo can be computed as (similar to Eq. \eqref{ZEqnNum260973}), 
\begin{equation} 
\label{ZEqnNum417020} 
\begin{split}
\Phi _{h} =-\gamma _{\Phi } \frac{GM}{r_{h} } &=-\frac{c\left\{1-{\left[1+2\left(1+c\right)\ln \left(1+c\right)\right]/\left(1+c\right)^{2} } \right\}}{2\left[\ln \left(1+c\right)-{c/\left(1+c\right)} \right]^{2} } \frac{GM}{r_{h} }\\ &=-\left(0.96\sim 1.26\right)\frac{GM}{r_{h} }.
\end{split}
\end{equation} 

The one-dimensional velocity dispersion $\sigma _{v}^{2} $ can be similarly obtained by the virial theorem,
\begin{equation} 
\label{ZEqnNum710824} 
\begin{split}
\sigma _{v}^{2}&=\gamma _{v} \frac{c\left\{1-{\left[1+2\left(1+c\right)\ln \left(1+c\right)\right]/\left(1+c\right)^{2} } \right\}}{6\left[\ln \left(1+c\right)-{c/\left(1+c\right)} \right]^{2} } \frac{GM}{r_{h} }\\ &=\left(0.32\sim 0.42\right)\gamma _{v} \frac{GM}{r_{h} }.
\end{split}
\end{equation} 
\end{enumerate}

In principle, if density profile is known for halos with mass $m_{h} $, all three quantities ($r_{g} $, $\Phi _{h} $, and $\sigma _{v}^{2} $) can be modeled by three constants $\gamma _{g} $, $\gamma _{v} $, $\gamma _{\Phi } $ and a critical density ratio $\Delta _{c} $, where
\begin{equation} 
\label{ZEqnNum785173} 
r_{g}^{} =\gamma _{g} r_{h} =\gamma _{g} a\left(\frac{2Gm_{h} }{\Delta _{c} H_{0}^{2} } \right)^{{1/3} } , \Phi _{h} =-\gamma _{\Phi } \frac{Gm_{h} }{r_{h} } =-\frac{\gamma _{\Phi } }{\gamma _{g} } \frac{Gm_{h} }{r_{g} }  
\end{equation} 
and
\begin{equation} 
\label{ZEqnNum676335} 
\sigma _{v}^{2} =-\Phi _{h} \frac{\gamma _{v} }{3} =\frac{1}{3} \gamma _{\Phi } \gamma _{v} \left(\frac{\Delta _{c} }{2} \right)^{{1/3} } \left(Gm_{h} H_{0} \right)^{{2/3} } a^{-1} .      
\end{equation} 
The critical density ratio $\Delta _{c} =18\pi ^{2} $ can be obtained from either a spherical collapse model or a two-body collapse model (TBCM) \citep[see][Eq. (89)]{Xu:2021-A-non-radial-two-body-collapse}.

All halos identified in \textit{N}-body simulation are first partitioned into groups of halos with the same mass $m_{h} $. In general, the halo kinetic energy $\sigma _{v}^{2} $ and root mean square radius $r_{g} $ in Eq. \eqref{ZEqnNum753014} can be dependent on halo mass \citep[see][Figs. 2 and 13]{Xu:2021-Inverse-and-direct-cascade-of-}. However, it is possible to introduce two new constants that are relatively independent of halo mass,
\begin{equation}
\beta _{s}^{*} =\frac{Hr_{g} }{\sigma _{v} } \quad \textrm{and} \quad \alpha _{s}^{*} =\frac{\sigma _{v}^{2} r_{g} }{Gm_{h} },       \label{ZEqnNum569238}
\end{equation}

\noindent where the root mean square radius $r_{g} $ and velocity dispersion $\sigma _{v} $ can be easily computed for each halo using Eq. \eqref{ZEqnNum753014}. We compute  two dimensionless parameters $\alpha _{s}^{*} $ and $\beta _{s}^{*} $ for every halo in the same group followed by averaging performed over all halos from the same group, where $\left\langle \bullet \right\rangle$ and 'std' stands for the mean value and the standard deviation of quantity `$\bullet $' for that halo group. 

By substituting Eqs. \eqref{ZEqnNum785173} and \eqref{ZEqnNum676335} into Eq. \eqref{ZEqnNum569238}, two constants are related to $\gamma _{g} $, $\gamma _{v} $, $\gamma _{\Phi } $ and $\Delta _{c} $ as
\begin{equation}
\beta _{s}^{*} =\frac{Hr_{g} }{\sigma _{v} } =\sqrt{\frac{6\gamma _{g}^{2} }{\gamma _{\Phi } \gamma _{v} \Delta _{c} } } \quad \textrm{and} \quad \alpha _{s}^{*} =\frac{\sigma _{v}^{2} r_{g} }{Gm_{h} } =\frac{1}{3} \gamma _{\Phi } \gamma _{v} \gamma _{g}.    
\label{ZEqnNum459478}
\end{equation}

\noindent With $\Delta _{c} =18\pi ^{2} $, $\gamma _{\Phi } =1$, ${\gamma _{g} =\sqrt{3}/3} $, and $\gamma _{v} =1.5$ for isothermal density profile \citep[see][Table 3]{Xu:2022-The-mean-flow--velocity-disper}, we expect $\beta _{s}^{*} \approx 0.09$ and $\alpha _{s}^{*} \approx 0.29$ for large halos with $m_{h} \to \infty $. The constant $\beta _{s}^{*} $ quantifies the competition between expanding background (via Hubble parameter \textit{H}) and gravity and is closely related to the critical value $\beta _{s2}$ in two-body collapse model \citep[see][Eq. (82)]{Xu:2021-A-non-radial-two-body-collapse}. To form an equilibrium collapse for halos with infinitesimally small lifetime, $\beta _{s2} ={1/\left(3\pi \right)}$ such that $\beta _{s}^{*} =\sqrt{2} \gamma _{g} \beta _{s2} =\sqrt{{2/3} } \beta _{s2} \approx 0.087$ (from Eq. \eqref{ZEqnNum459478}). 

\begin{figure}
\includegraphics*[width=\columnwidth]{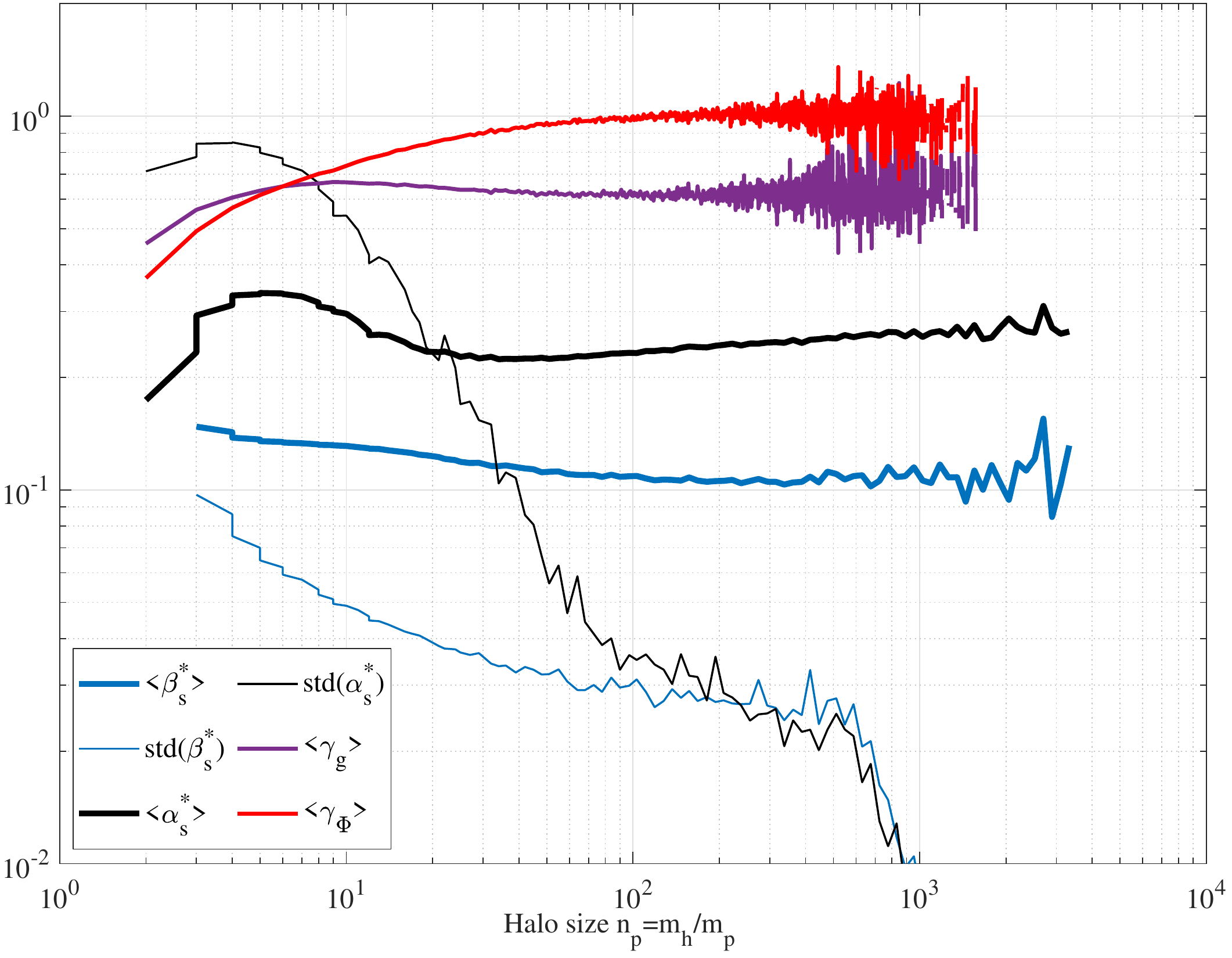}
\caption{The variation of mean and standard deviation of two dimensionless parameters $\alpha _{s}^{*} $ and $\beta _{s}^{*} $ with halo size $n_{p} $ from a \textit{N}-body simulation at \textit{z}=0 ($m_{p} =2.27\times 10^{11} {M_{\odot }/h} $). The dispersion of $\alpha _{s}^{*} $ and $\beta _{s}^{*} $ (the standard deviation) decreases with halo size because of the decreasing halo lifetime with halo size and halo synchronization. The limiting values of $\alpha _{s}^{*} $ and $\beta _{s}^{*} $ for large halos approach 0.29 and 0.09 (Eq. \eqref{ZEqnNum459478}), respectively. The variation of $\gamma _{g} $ and $\gamma _{\Phi } $ is also presented in the same plot using Eq. \eqref{ZEqnNum561187}.}
\label{fig:2}
\end{figure}

Figure \ref{fig:2} plots the variation of mean and standard deviation of $\alpha _{s}^{*} $ and $\beta _{s}^{*} $ with halo mass $m_{h} $ (or equivalently the halo size $n_{p} $) at \textit{z}=0. As expected, both $\alpha _{s}^{*} $ and $\beta _{s}^{*} $ have much wider distributions (larger standard deviation) for small halos and converge to constant values for large halos. Small halos have longer lifetime when compared to large halos with a scaling of lifetime with halo mass as $\tau _{g} \sim m_{h}^{{}^{{-2/3} } } $ \citep[see][Eq. (45)]{Xu:2021-Inverse-mass-cascade-mass-function}. At a given redshift \textit{z}, small halos of same size can be generated at different time while large halos tend to be synchronized and generated at the same time. Therefore, small halos have a wider distribution of properties while large halos have similar properties with a narrower distribution. Both constants approach limiting values for large halos that can be very well estimated by Eq. \eqref{ZEqnNum459478}, i.e. $\beta _{s}^{*} \approx 0.09$ and $\alpha _{s}^{*} \approx 0.29$. 

With $\alpha _{s}^{*} $ and $\beta _{s}^{*}$ (Fig. \ref{fig:2}) and $\gamma _{v} =-n_{s}^{*} $ (in Fig. \ref{fig:5}) determined for each halo in simulation, and critical density ratio $\Delta _{c} =18\pi ^{2}$, other two constants for each halo group can be determined using Eq. \eqref{ZEqnNum459478}, 
\begin{equation}
\gamma _{g} =\left(\frac{1}{2} \alpha _{s}^{*} \beta _{s}^{*2} \Delta _{c} \right)^{{1/3} } \quad \textrm{and} \quad \gamma _{\Phi } =\frac{6\left({\alpha _{s}^{*} \beta _{s}^{*2} \Delta _{c}/2} \right)^{{2/3} } }{\gamma _{v} \beta _{s}^{*2} \Delta _{c}}.    \label{ZEqnNum561187}
\end{equation}

The variation of $\gamma _{g} $ and $\gamma _{\Phi } $ with halo size is also presented in Fig. \ref{fig:2}, with $\gamma _{g} $ slightly decreasing with halo size and $\gamma _{\Phi } $ increasing with halo size and approaching 1. Now with dimensionless constants $\gamma _{g} $, $\gamma _{v} $, $\gamma _{\Phi } $ determined, the root mean square radius $r_{g} $, specific potential $\Phi _{h} $, and virial dispersion $\sigma _{v}^{2} $ can be easily computed for halos with known mass $m_{h}$ (Eqs. \eqref{ZEqnNum785173} and \eqref{ZEqnNum676335}).

\subsection{Halo radial and angular momentum, and spin parameter}
\label{sec:4.3}
The temporal evolution of virial quantity (radial momentum) and angular momentum in a transformed system is discussed in Section \ref{sec:3.3}. We can define them in a comoving system accordingly,
\begin{equation}
\boldsymbol{\mathrm{H}}_{p} =\frac{1}{N} \sum _{i=1}^{N}\boldsymbol{\mathrm{x}}_{i} \times \boldsymbol{\mathrm{u}}_{i} \quad \textrm{and} \quad G_{p} =\frac{1}{N} \sum _{i=1}^{N}\boldsymbol{\mathrm{x}}_{i} \cdot \boldsymbol{\mathrm{u}}_{i}, \label{ZEqnNum512486}
\end{equation}
where the summation runs over all \textit{N} particles in N-body system. 

Next, all particles in entire \textit{N}-body system can be decomposed into a halo sub-system and an out-of-halo sub-system. The halo sub-system includes all particles that reside in all halos identified, while the out-of-halo sub-system includes all particles that do not belong to any halos. The same quantity can be computed equivalently for halo sub-system over all halos ($N_{h} $) and all particles ($n_{p} $) in every halo. For every halo, we first decompose the particle motion into the motion of halo and the motion in halo,
\begin{equation}
\begin{split}
&\boldsymbol{\mathrm{x}}_{i} =\boldsymbol{\mathrm{x}}_{h} +\boldsymbol{\mathrm{x}}_{i}^{'} =\left(\frac{1}{n_{p} } \sum _{i=1}^{n_{p} }\boldsymbol{\mathrm{x}}_{i}  \right)+\boldsymbol{\mathrm{x}}_{i}^{'}\\ 
&\textrm{and}\\
&\boldsymbol{\mathrm{u}}_{i} =\boldsymbol{\mathrm{u}}_{h} +\boldsymbol{\mathrm{u}}_{i}^{'} =\left(\frac{1}{n_{p} } \sum _{i=1}^{n_{p} }\boldsymbol{\mathrm{u}}_{i}  \right)+\boldsymbol{\mathrm{u}}_{i}^{'},
\end{split}
\label{ZEqnNum561926}
\end{equation}

\noindent where $\boldsymbol{\mathrm{x}}_{h} $ is the position of the center of halo mass and $\boldsymbol{\mathrm{u}}_{h} $ is the halo peculiar velocity. Here $\boldsymbol{\mathrm{x}}_{i}^{'} $ and $\boldsymbol{\mathrm{u}}_{i}^{'} $ are the relative position and velocity of particles in that halo. By using Eq. \eqref{ZEqnNum561926}, we can decompose the momentum of  halo sub-system $\boldsymbol{\mathrm{H}}_{hs} $ and $G_{hs} $ into contributions from motion in halo and motion of halo,  
\begin{equation}
\begin{split}
&\boldsymbol{\mathrm{H}}_{hs} =\frac{1}{N_{hp} } \sum _{j=1}^{N_{h} }n_{p} \left(\boldsymbol{\mathrm{H}}_{hc} +\boldsymbol{\mathrm{H}}_{ec} \right)\\
&G_{hs} =\frac{1}{N_{hp} } \sum _{j=1}^{N_{h} }n_{p} \left(G_{hc} +G_{ec} \right) \quad \textrm{and} \quad N_{hp} =\sum _{j=1}^{N_{h} }n_{p},
\end{split}
\label{ZEqnNum727626}
\end{equation}

\noindent where $N_{hp} $ is the total number of particles in all halos. The subscript `\textit{h'} stands for that quantity from the motion of particle in halos and `\textit{c}' stands for the comoving coordinate, 
\begin{equation}
\boldsymbol{\mathrm{H}}_{hc} =\frac{1}{n_{p} } \sum _{i=1}^{n_{p} }\left(\boldsymbol{\mathrm{x}}_{p}^{'} \times \boldsymbol{\mathrm{u}}_{p}^{'} \right) \quad \textrm{and} \quad G_{hc} =\frac{1}{n_{p} } \sum _{i=1}^{n_{p} }\left(\boldsymbol{\mathrm{x}}_{p}^{'} \cdot \boldsymbol{\mathrm{u}}_{p}^{'} \right).    
\label{ZEqnNum838771}
\end{equation}

\noindent The subscript \textit{e} stands for that quantity from motion of entire halo,
\begin{equation}
\boldsymbol{\mathrm{H}}_{ec}^{} =\boldsymbol{\mathrm{x}}_{h} \times \boldsymbol{\mathrm{u}}_{h} \quad \textrm{and} \quad G_{ec}^{} =\boldsymbol{\mathrm{x}}_{h} \cdot \boldsymbol{\mathrm{u}}_{h}.     \label{eq:60}
\end{equation}

If all particles in N-body system reside in halos and there are no out-of-halo particles, then $G_{hs} =G_{p} $ and $\boldsymbol{\mathrm{H}}_{hs} =\boldsymbol{\mathrm{H}}_{p}$ in Eq. \eqref{ZEqnNum512486}. The virial quantity and angular momentum for entire system can also be computed in physical coordinate and denoted as $G_{py} $ and $\boldsymbol{\mathrm{H}}_{py}$ (just like $G_{p}$ and $\boldsymbol{\mathrm{H}}_{p}$ in comoving coordinate in Eq. \eqref{ZEqnNum512486}). For halos, $G_{h} =aG_{hc}$ and $\boldsymbol{\mathrm{H}}_{h} =a\boldsymbol{\mathrm{H}}_{hc}$ are the virial quantity and angular momentum in physical coordinate (just like $G_{hc}$ and $\boldsymbol{\mathrm{H}}_{hc}$ in comoving coordinates in Eq. \eqref{ZEqnNum838771}). 

\begin{figure}
\includegraphics*[width=\columnwidth]{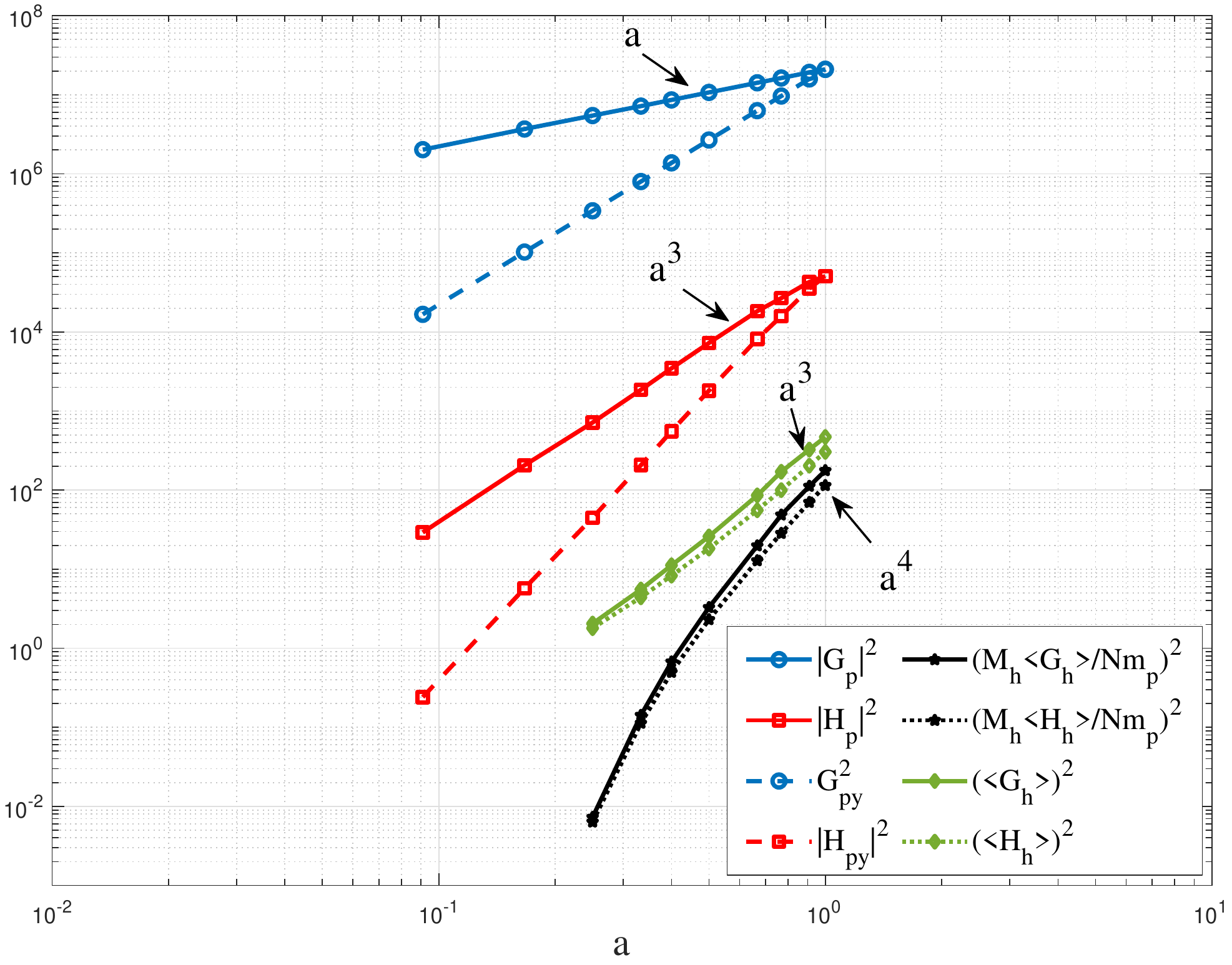}
\caption{The time evolution of virial quantity and angular momentum ($(km/s\cdot Mpc/h)^2$) of entire system and halo sub-system with scale factor \textit{a} from \textit{N}-body simulation. All quantities exhibit a power-law scaling with scale factor \textit{a}. An approximate scaling of $G_{p} \left(a\right)\propto a^{{1/2} } $ and $H_{p} \left(a\right)\propto a^{{3/2} } $ can be identified for two comoving quantities. The same quantities $G_{py} \propto a^{{3/2} } $ and $H_{py} \propto a^{{5/2} } $ (in physical coordinates) for entire system are also plotted. The evolution of halo-averaged specific quantities (momentum in physical coordinate averaged over all halo particles) $\left\langle G_{h} \right\rangle \propto a^{{3/2} } $ and $\left\langle \left|\boldsymbol{\mathrm{H}}_{h} \right|\right\rangle \propto a^{{3/2}}$ are presented in green lines. The total virial quantity and angular momentum in all halos scales as $\sim a^{2} $ with total halo mass $M_{h} \sim a^{{1/2} } $ (black lines).}
\label{fig:3}
\end{figure}

Figure \ref{fig:3} presents the time variation of the specific virial quantity $G_{p} $ and angular momentum $H_{p} =\left|\boldsymbol{\mathrm{H}}_{p} \right|$ (in comoving coordinate) from \textit{N}-body simulation in Section \ref{sec:2}. Both quantities exhibit a power-law scaling with scale factor \textit{a}, with an approximate scaling of $G_{p} \left(a\right)\propto a^{{1/2} } $ and $H_{p} \left(a\right)\propto a^{{3/2} } $.  The same quantities $G_{py} $ and $\boldsymbol{\mathrm{H}}_{py} $ in physical coordinate are also presented in the same figure with $G_{py} \propto a^{{3/2} } $ and $\left|\boldsymbol{\mathrm{H}}_{py} \right|\propto a^{{5/2} } $. The halo averaged specific quantities $\left\langle G_{h} \right\rangle $ and $\left\langle \left|\boldsymbol{\mathrm{H}}_{h} \right|\right\rangle $ (average over all particles in all halos) in physical coordinate are also presented with the scaling of $\left\langle G_{h} \right\rangle \sim a^{{3/2} } \sim t$ and $\left\langle \left|\boldsymbol{\mathrm{H}}_{h} \right|\right\rangle \sim a^{{3/2} } \sim t$ (green lines). This is consistent with the momentum results for typical halos ("large" halos at its early stage with fast mass accretion) that grow with a constant waiting time to merge with single merger $\tau_g$ \citep[see][Eq. (45)]{Xu:2021-Inverse-mass-cascade-mass-function} \citep[see][Table 3]{Xu:2022-The-mean-flow--velocity-disper}. The total virial quantity and angular momentum in all halos scale as $M_{h} \left\langle G_{h} \right\rangle $$\sim a^{2} $ and $M_{h} \left\langle \left|\boldsymbol{\mathrm{H}}_{h} \right|\right\rangle \sim a^{2} $ with total halo mass $M_{h} \sim a^{{1/2} } $ (black lines). 

\begin{figure}
\includegraphics*[width=\columnwidth]{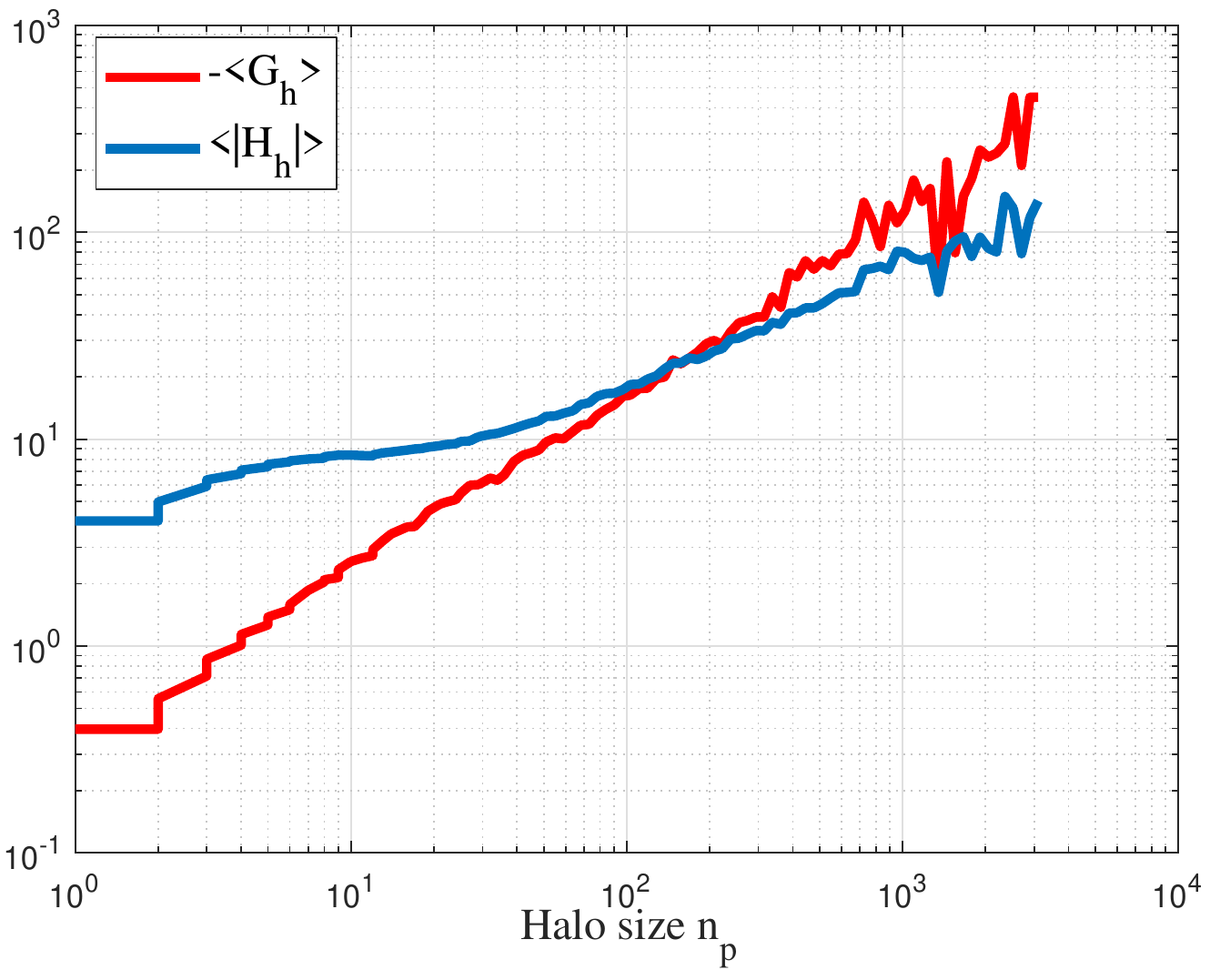}
\caption{The variation of halo virial quantity and angular momentum $(km/s\cdot Mpc/h)$ with halo size $n_{p} $ at \textit{z}=0. Halo momentum quantities $G_{h} $ and $\left|\boldsymbol{\mathrm{H}}_{h} \right|$ are due to the particle radial and rotational motion in halos. The rotational motion $\left|\boldsymbol{\mathrm{H}}_{h} \right|$ is dominant over the radial motion $G_{h} $ for rotation supported (small) halos ($\left|\boldsymbol{\mathrm{H}}_{h} \right|\approx -3\pi G_{h} a^{{3/2} } $ from \citep{Xu:2021-A-non-radial-two-body-collapse}), while they are comparable for large halos.}
\label{fig:4}
\end{figure}

The halo virial quantity $G_{h}$ and angular momentum $\boldsymbol{\mathrm{H}}_{h}$ in physical coordinate can be easily computed for every halo identified in the simulation, which involves the dot and cross products between $\boldsymbol{\mathrm{u}}_{p}^{'} $ and $\boldsymbol{\mathrm{x}}_{p}^{'} $ (Eq. \eqref{ZEqnNum838771}). Here $G_{h} $ and $\boldsymbol{\mathrm{H}}_{h} $ quantify particle motion in radial and tangential directions, respectively. The halo-group averaged virial quantity $G_{h}$ and angular momentum $\left|\boldsymbol{\mathrm{H}}_{h} \right|$ for all halos of the same size $n_{p} $ in the same halo group are presented in Fig. \ref{fig:4}.

Both halo momentum quantities increase with halo size $n_{p} $ with $G_{h} <0$. Particles in outer region of halo tend to fall inward due to gravitational interaction. Rotational motion $\left|\boldsymbol{\mathrm{H}}_{h} \right|$ is dominant over the radial motion $G_{h} $ for small halos, where $\left|\boldsymbol{\mathrm{H}}_{h} \right|\approx -3\pi G_{h} a^{{3/2} } $ from a two-body collapse model (TBCM model, \citep[see Eq. (104)]{Xu:2021-A-non-radial-two-body-collapse}). The factor $3\pi$ comes from the critical value $\beta_{s2}=1/(3\pi)$ for the ratio of circular velocity to radial velocity. Two quantities are comparable for large halos.

Halo virial quantity (radial momentum) and angular momentum in physical coordinate (Fig. \ref{fig:4}) have been modelled in our previous work \citep{Xu:2021-Inverse-and-direct-cascade-of-},   
\begin{equation}
G_{h} \approx -f_{G} \left(m_{h} \right)a^{-1} Hr_{g}^{2} \quad \textrm{and} \quad \left|\boldsymbol{\mathrm{H}}_{h} \right|\approx f_{H} \left(m_{h} \right)a^{{1/2} } Hr_{g}^{2},      
\label{ZEqnNum483595}
\end{equation}
where two dimensionless functions $f_{G} \left(m_{h} \right)$ and $f_{H} \left(m_{h} \right)$ are presented in \citep[see][Figs. 12 and 14]{Xu:2021-Inverse-and-direct-cascade-of-}. 

Here with root mean square radius $r_{g} $ and virial dispersion $\sigma _{v}^{2} $ (Eqs. \eqref{ZEqnNum785173} and \eqref{ZEqnNum676335}), two proportional coefficients $\tau _{s}^{*} $ and $\eta _{s}^{*} $ can be used to conveniently relate $G_{h} $ and $\left|\boldsymbol{\mathrm{H}}_{h} \right|$ with the typical halo size $r_{g} $ and typical velocity $\sigma _{v}^{2} $ for each halo,

\begin{equation}
G_{h} =-\tau _{s}^{*} \sigma _{v} r_{g} a^{-1} \quad \textrm{and} \quad \left|\boldsymbol{\mathrm{H}}_{h} \right|=\eta _{s}^{*} \sigma _{v} r_{g} a^{{1/2} },     
\label{ZEqnNum336536}
\end{equation}

\noindent where both coefficients $\tau _{s}^{*} $ and $\eta _{s}^{*} $ can be functions of halo mass $m_{h} $. With Eqs. \eqref{ZEqnNum785173} and \eqref{ZEqnNum676335}, coefficients $\tau _{s}^{*} $ and $\eta _{s}^{*} $ are related to the constant $\beta _{s}^{*} $ and two functions $f_{G} \left(m_{h} \right)$ and $f_{H} \left(m_{h} \right)$ (defined in Eqs. \eqref{ZEqnNum569238} and \eqref{ZEqnNum459478}) as,
\begin{equation}
\tau _{s}^{*} =\beta _{s}^{*} f_{G} \left(m_{h} \right) \quad \textrm{and} \quad \eta _{s}^{*} =\beta _{s}^{*} f_{H} \left(m_{h} \right)     
\label{ZEqnNum823621}
\end{equation}

\noindent where $\tau _{s}^{*} $ and $\eta _{s}^{*} $ are dimensionless measures of halo radial ($G_{h} $) and angular momentum ($\left|\boldsymbol{\mathrm{H}}_{h} \right|$). 

Equations \eqref{ZEqnNum483595} and \eqref{ZEqnNum336536} are not empirical and can be interpreted as follows: for halos with a given mass $m_{h} $, the angular momentum is expected to grow as $\boldsymbol{\mathrm{H}}_{h} \propto r_{g} \propto a$ (Eqs. \eqref{ZEqnNum785173} and \eqref{ZEqnNum336536}), such that the angular momentum of two-body halos can be exactly written as,
\begin{equation} 
\label{ZEqnNum292188} 
\left|\boldsymbol{\mathrm{H}}_{h} \right|=\left(\sqrt{3} \sigma _{v} a^{{1/2} } \right)\left({r_{g}/a} \right)a,
\end{equation} 
where $\sigma _{v} \propto a^{{-1/2} } $ and $r_{g} \propto a$ in Eqs. \eqref{ZEqnNum785173} and \eqref{ZEqnNum676335} for two-body halos, and $\sqrt{3} \sigma _{v} a^{{1/2}} $ is the particle velocity in transformed system (Eq. \eqref{ZEqnNum815118}). Note that the factor $\sqrt{3}$ comes from the fact that $\sigma _{v}^{2} $ is one-dimensional dispersion and ${r_{g}/a} $ is the comoving separation between two particles. Note that $\eta _{s}^{*} =\sqrt{3}$ for two-body halos by comparing Eqs. \eqref{ZEqnNum292188} and \eqref{ZEqnNum336536}. 

From a TBCM model, the dimensionless constant that quantifies the competition between expanding background and gravity is defined as \citep[see][Eq. (82)]{Xu:2021-A-non-radial-two-body-collapse} with a critical value ${1/(3\pi)}$,
\begin{equation} 
\label{ZEqnNum678401} 
\beta _{s} =\frac{H_{0} \left({r_{g}/a} \right)}{\left(\sqrt{3} \sigma _{v} a^{{1/2} } \right)} =\beta _{s2} =\frac{1}{3\pi } .         
\end{equation} 

The virial quantity (radial momentum) and angular momentum of two-body halos can be finally written as (using Eqs. \eqref{ZEqnNum292188}, \eqref{ZEqnNum678401} and Eq. (104) from \cite{Xu:2021-A-non-radial-two-body-collapse}),
\begin{equation}
\begin{split}
&|\boldsymbol{\mathrm{H}}_{h}|=3\pi H_{0} \left({r_{g}/a} \right)^{2} a=3\pi a^{{1/2} } Hr_{g}^{2}\\
&\textrm{and}\\
&G_{h} =-{\left|\boldsymbol{\mathrm{H}}_{h} \right|a^{-{3/2} }/(3\pi)} =-a^{-1} Hr_{g}^{2},  
\end{split}
\label{ZEqnNum112040}
\end{equation}
such that $f_{H} \left(m_{h} \to 0\right)=3\pi $ and $f_{G} \left(m_{h} \to 0\right)=1$ by comparing with Eq. \eqref{ZEqnNum483595}. 

The expression for two-body halos in Eq. \eqref{ZEqnNum292188} can be generalized to halos of any size in Eq. \eqref{ZEqnNum336536}, while the coefficient $\eta _{s}^{*} $ is expected to decrease with halo mass. For large halos, particles in the same spherical shell have similar speed on the order of $\sim \sigma _{v} $ but along different directions. The contribution from all particles in the same spherical shell to halo angular momentum is partially cancelled out due to particle velocity along random directions. This explains a smaller $\eta _{s}^{*}$ for larger halos in Fig. \ref{fig:6}. 

Finally, the halo angular momentum is usually described by a dimensionless spin parameter \citep{Efstathiou:1979-Rotation-of-Galaxies---Numeric}
\begin{equation} 
\label{eq:67} 
\lambda _{p} =\frac{\left|\boldsymbol{\mathrm{H}}_{h} \right|\left|E_{h} \right|^{{1/2} } }{Gm_{h} } ,          
\end{equation} 
which represents the ratio between rotational energy and total energy of that halo. Here $E_{h} =K_{h} +\Phi _{h} $ is the total halo specific energy. The halo spin parameter characterizes the importance of rotational motion to random motion. Next, halo spin $\lambda _{p} $ can be related to $\eta _{s}^{*} $ by introducing parameter $z_{s}^{*}$ and a halo effective exponent $n_{s}^{*} $ defined as (similar to $n_{e} $ for the entire N-body system in Eq. \eqref{ZEqnNum539248}),  
\begin{equation}
z_{s}^{*} =\frac{E_{h} }{\sigma _{v}^{2} } =\frac{K_{h} +\Phi _{h} }{\sigma _{v}^{2} } \quad \textrm{and} \quad n_{s}^{*} =\frac{2K_{h} }{\Phi _{h} } =\frac{3\sigma _{v}^{2} }{\Phi _{h} } =-\gamma _{v}.     
\label{ZEqnNum400737}
\end{equation}

\noindent Clearly, $z_{s}^{*} $ and $n_{s}^{*} $ are related by
\begin{equation} 
\label{eq:69} 
z_{s}^{*} =\frac{3}{2} +\frac{3}{n_{s}^{*} }  .           
\end{equation} 

\begin{figure}
\includegraphics*[width=\columnwidth]{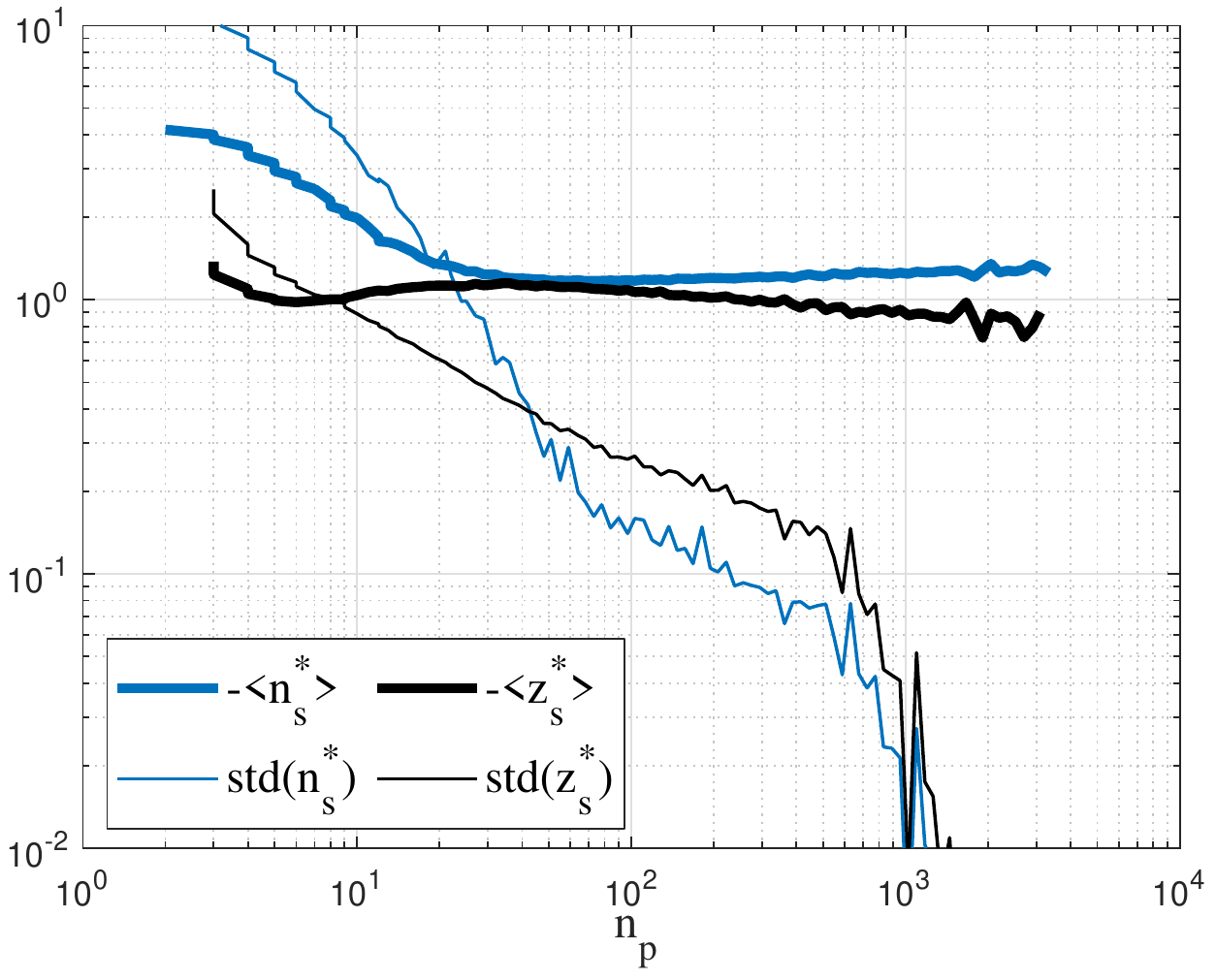}
\caption{The variation of mean and standard deviation of halo parameters $n_{s}^{*} $ and $z_{s}^{*} $ with halo size $n_{p} $ from \textit{N}-body simulation at \textit{z}=0. The standard deviation of $n_{s}^{*} $ and $z_{s}^{*} $ for halo groups decreases with the halo size because of short lifetime and synchronization of large halos. The halo ratio parameter $z_{s}^{*} \approx -0.81$ and the halo potential exponent $n_{s}^{*} \approx -1.3\ne -1$ for large halos due to halo mass accretion and halo surface energy \citep[see][Eq. (96)]{Xu:2021-Inverse-mass-cascade-halo-density}.}
\label{fig:5}
\end{figure}

We compute the mean and standard deviation of two halo parameters for groups of halos of the same mass $m_{h} $ (or equivalently same $n_{p} $, the number of particles in a halo). Figure \ref{fig:5} plots the variation of mean and the standard deviation of $n_{s}^{*} $ and $z_{s}^{*} $ from a \textit{N}-body simulation at \textit{z}=0. The standard deviation of $n_{s}^{*} $ and $z_{s}^{*} $ decreases with halo size $n_{p} $ because of short lifetime and synchronization of large halos. The halo ratio parameter $z_{s}^{*} \approx -0.81$, while halo potential exponent $n_{s}^{*} \approx -1.3\ne -1$ for large halos \citep[see][Eq. (96)]{Xu:2021-Inverse-mass-cascade-halo-density}, which agrees with the effective potential exponent $n_{e} $ for entire system in Fig. \ref{fig:1b} when dynamics in halos becomes dominant. 

With Eqs. \eqref{ZEqnNum483595} and \eqref{ZEqnNum400737}, the halo spin parameter $\lambda _{p} $ for a given halo size is finally written as
\begin{equation} 
\label{ZEqnNum879931} 
\begin{split}
\lambda _{p} =a^{{1/2} } \alpha _{s}^{*} \eta _{s}^{*} \sqrt{\left|z_{s}^{*} \right|}&=a^{{1/2} } \alpha _{s}^{*} \beta _{s}^{*} \sqrt{\left|z_{s}^{*} \right|} f_{H} \left(m_{h} \right)\\
&=a^{{1/2} } \gamma _{g}^{2} f_{H} \left(m_{h} \right)\sqrt{\frac{2\gamma _{\Phi } \gamma _{v} \left|z_{s}^{*} \right|}{3\Delta _{c} } }.
\end{split}
\end{equation} 
With constants $\alpha _{s}^{*} $, $\beta _{s}^{*} $, and $z_{s}^{*} $ (in Figs. \ref{fig:2} and \ref{fig:5}) are all relatively independent of halo mass $m_{h} $, the mass dependence of spin parameter $\lambda _{p} $ mostly comes from function $f_{H} \left(m_{h} \right)$ or $\eta _{s}^{*} $. It can be shown that halo spin parameter $\lambda _{p} $ decreases with halo mass in \ref{fig:6}, which is consistent with $\eta _{s}^{*} $ in Fig. \ref{fig:6} and $f_{H} (m_{h})$ \citep[see][Fig. 14]{Xu:2021-Inverse-and-direct-cascade-of-}. 

For small halos with extremely slow mass accretion (or both $m_h$ and $f_{H}(m_{h})$ are independent of time), halo spin $\lambda _{p}$ increases as $\propto a^{{1/2}}$. This result for halo spin parameter agrees with other simulations \citep{Hetznecker:2006-The-evolution-of-the-dark-halo}. For large halos with extremely fast mass accretion, the spin parameter $\lambda _{p} $ is relatively a constant with $\lambda _{p} \approx 0.031$ \citep[also see][Eq. (119) and Fig. 14]{Xu:2022-The-mean-flow--velocity-disper}. This is because $f_{H}(m_{h})\propto m_{h}^{-{1/3}}$ for large halos with halo mass $m_{h} \propto a^{{3/2}}$.

The spin parameter $\lambda _{p}$ for two-body halos can be easily obtained by a two-body collapse model (TBCM), where
\begin{equation}
\label{ZEqnNum885344} 
\lambda _{p} =\frac{\left|\boldsymbol{\mathrm{H}}_{s} \right|\left|E_{s} \right|^{{1/2} } }{G\left(m_{1} +m_{2} \right)} =\frac{\sqrt{2} }{2} \frac{\left(m_{1} m_{2} \right)^{{3/2} } }{\left(m_{1} +m_{2} \right)^{3} }  
\end{equation} 
with equations for $\boldsymbol{\mathrm{H}}_{s} $, $E_{s}$ \citep[see][Eqs. (100), (94), (24)]{Xu:2021-A-non-radial-two-body-collapse} , 
\begin{equation} 
\label{eq:72} 
E_{s} =-\frac{2m_{1} m_{2}^{} v_{i}^{2} }{\left(m_{1} +m_{2} \right)^{2} } \exp \left(H_{0} s\right),         
\end{equation} 
\begin{equation}
\left|\boldsymbol{\mathrm{H}}_{s} \right|=\frac{4m_{1} m_{2} v_{i} r_{i}^{} }{\left(m_{1} +m_{2} \right)^{2} } \exp \left(-\frac{H_{0}s}{2}\right) \quad \textrm{and} \quad v_{i}^{2} = \frac{G\left(m_{1} +m_{2} \right)}{8r_{i}^{}}.   
\label{eq:73}
\end{equation}

\noindent The spin parameter $\lambda _{p} ={\sqrt{2}/16} \approx 0.0884$ can be obtained from Eq. \eqref{ZEqnNum885344} for two-body halos with equal masses $m_{1} =m_{2} $, i.e. most two-body halos (the mode of probability distribution of $\lambda _{p}$ of all two-body halos) should have a spin parameter $\lambda _{p} \approx 0.09$ that is much larger than $\lambda _{p} \approx 0.031$ for large halos.

\begin{figure}
\includegraphics*[width=\columnwidth]{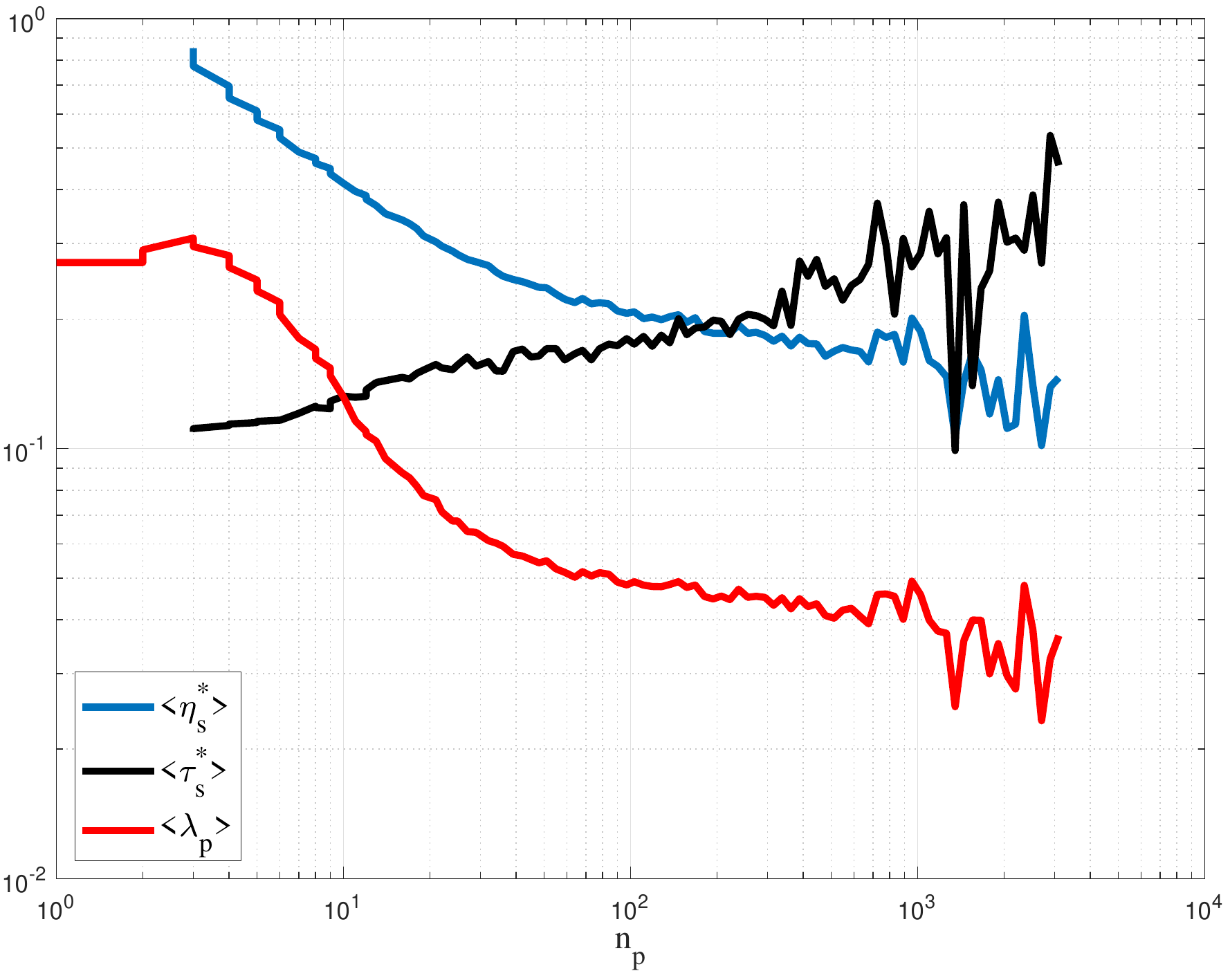}
\caption{The variation of mean halo parameters $\eta _{s}^{*} $, $\tau _{s}^{*} $, and halo spin parameter $\lambda _{p} $ with halo size $n_{p} $ ($m_{p} =2.27\times 10^{11} {M_{\odot }/h} $) from a \textit{N}-body simulation at \textit{z}=0. Parameters $\eta _{s}^{*} \gg \tau _{s}^{*} $ for small halos that are mostly rotation supported, while $\eta _{s}^{*} <\tau _{s}^{*} $ for large halos. The spin parameter $\lambda _{p} $ approaches 0.03$\mathrm{\sim}$0.04 for large halos and 0.3 for small halos.}
\label{fig:6}
\end{figure}

Figure \ref{fig:6} plots the variation of mean halo parameters $\eta _{s}^{*} $ and $\tau _{s}^{*} $ in Eq. \eqref{ZEqnNum336536} and the halo spin parameter $\lambda _{p} $ (Eq. \eqref{ZEqnNum879931}) with halo size $n_{p} $ from a \textit{N}-body simulation at \textit{z}=0. Parameter $\eta _{s}^{*} \gg \tau _{s}^{*} $ for small halos that are mostly rotation supported and $\eta _{s}^{*} <\tau _{s}^{*} $ for large halos. Especially for halo spin parameter, it is demonstrated that $\lambda _{p} \approx 0.28$ for small halos (mean $\lambda_p$ for all two-body halos), while $\lambda _{p} \approx 0.03$ for large halos that can be directly related to the critical value $\beta _{s2} $ (Eq. \eqref{ZEqnNum879931}). With $\gamma _{v} $, $\gamma _{g} $, $\gamma _{\Phi } $ and $\Delta _{c} $ determined, other relevant parameters can be easily calculated. Table \ref{tab:3} summarizes all relevant parameters for halo energy, momentum, and spin parameter from theory and simulation for both small and large halos.

\begin{table}
\begin{tabular}{p{0.85in}p{0.6in}p{0.6in}p{0.6in}} 
\hline 
Type of halos&Two-body halos &Large halos\newline (NFW)  & Large halos\newline (Isothermal)\\
\hline
$\gamma _{\Phi }$ (Eq.\eqref{ZEqnNum417020})  &  1/4 & 0.936 &1 \\
\hline
$\gamma _{g}$ (Eq.\eqref{ZEqnNum159016}) & 1/2 & 0.567  & ${\sqrt{3}/3}$\\
\hline
$\gamma _{v}$ (Eq.\eqref{ZEqnNum676335}) & 1.0 & 1.3 & 1.5 \\
\hline
$\Delta _{c} $ &$18\pi ^{2} $ & $18\pi ^{2} $&$18\pi ^{2} $ \\
\hline
$\alpha _{s}^{*} $ (Eq.\eqref{ZEqnNum459478}) & 1/24 & 0.230 & ${\sqrt{3}/6} $\\
\hline
$\beta _{s}^{*} $ (Eq.\eqref{ZEqnNum459478})  & ${\sqrt{3}/\left(3\pi \right)} $  & 0.095 & ${\sqrt{{2 /3} }}/{\left(3\pi \right)}$\\
\hline
$n_{s}^{*} $ (Eq.\eqref{ZEqnNum400737}) & -1.0 & -1.3 & -1.5 \\
\hline
$z_{s}^{*} $ (Eq.\eqref{ZEqnNum400737}) & -1.5 & -0.81 & -0.5 \\
\hline
$\eta _{s}^{*} $ (Eq.\eqref{ZEqnNum823621}) & $\sqrt{3} $ & 0.151  & ${\sqrt{2/3}/\left(3\pi \right)} $ \\
\hline
$\tau _{s}^{*} $ (Eq.\eqref{ZEqnNum823621}) & ${\sqrt{3}/\left(3\pi \right)} $ & 0.103  & ${\sqrt{{2/3} }/\left(3\pi \right)} $ \\
\hline
$f_{H} \left(m_{h} \right)$ (Eq.\eqref{ZEqnNum823621}) & $3\pi$ &1.59 & 1 \\
\hline 
$f_{G} \left(m_{h} \right)$ (Eq.\eqref{ZEqnNum823621}) &1 &1.08  & 1 \\
\hline
$\lambda_{p}$ (Eq.\eqref{ZEqnNum879931}) & $\sqrt{2}/16$ & 0.031 & ${1/\left(18\pi \right)} $ \\
\hline
\end{tabular}
\caption{Relevant parameters for halo energy, momentum and spin from theory and simulations}
\label{tab:3}
\end{table}

\section{Integral constants of motion in dark matter flow}
\label{sec:5}
The virial quantity and angular momentum defined in Eqs. \eqref{ZEqnNum446862} and \eqref{ZEqnNum346758} are intimately related to the integral constants of motion for self-gravitating collisionless dark matter flow. The evolution of virial quantity and angular momentum are modelled in this section on large and halo scales (Eqs. \eqref{ZEqnNum374769}, \eqref{ZEqnNum748159}, and \eqref{ZEqnNum738790}).

\subsection{Integral constants of motion on large scale}
\label{sec:5.1}
Let's first define a series of constants $I_{m}$ for (\textit{m-2})th moment of velocity correlation function  
\begin{equation} 
\label{ZEqnNum530177} 
\begin{split}
I_{m} =\int \left\langle \boldsymbol{\mathrm{u}}\cdot \boldsymbol{\mathrm{u}}^{'} \right\rangle r^{m-2} d\boldsymbol{\mathrm{r}}^{3}&=\int R_{2} \left(r\right)r^{m-2} d\boldsymbol{\mathrm{r}}^{3}\\
&=\int _{0}^{\infty }4\pi R_{2} \left(r\right)r^{m} dr,
\end{split}
\end{equation} 
where $R_{2} \left(r\right)=\left\langle \boldsymbol{\mathrm{u}}\cdot \boldsymbol{\mathrm{u}}^{'} \right\rangle $ is the total velocity correlation for pair of particles at locations $\boldsymbol{\mathrm{x}}^{'} $ and $\boldsymbol{\mathrm{x}}$ with a distance $r=\left|\boldsymbol{\mathrm{r}}\right|=\left|\boldsymbol{\mathrm{x}}^{'} -\boldsymbol{\mathrm{x}}\right|$ and velocity $\boldsymbol{\mathrm{u}}^{'} $ and $\boldsymbol{\mathrm{u}}$ \citep[see] [Fig. 1 and Eq. (16)]{Xu:2022-The-statistical-theory-of-2nd} or \citep[see][Eq. (A8)]{Xu:2022-The-statistical-theory-of-3rd}. 

The second moment is $I_{2}$ (Saffman's integral) \citep{Saffman:1958-The-Penetration-of-a-Fluid} and the fourth moment is $I_{4}$ (Loitsyansky's integral) \citep{Loitsyansky:1939-Some-basic-laws-for-isotropic-}, both of which are relevant to the dynamics of turbulence on large scale. In dark matter flow, the energy spectrum is related to velocity correlation \citep[see Eq. (25)]{Xu:2022-The-statistical-theory-of-2nd},
\begin{equation} 
\label{eq:75} 
E_{u} \left(k\right)=\frac{1}{\pi } \int _{0}^{\infty }R_{2} \left(r\right)kr\sin \left(kr\right)dr .        
\end{equation} 
The Taylor expansion of term $\sin \left(kr\right)$ leads to the \textit{m}th order constant $I_{m} $ that is essentially the derivatives of the spectrum $E_{u} \left(k\right)$ around wavenumber $k=0$ (the long wavelength limit on large scale),
\begin{equation}
\label{ZEqnNum222875} 
I_{m} =4\pi ^{2} \frac{\left(-1\right)^{1+{m/2} } }{m} \left. \frac{\partial ^{m} E_{u} }{\partial k^{m} } \right|_{k=0} \propto a,        
\end{equation} 
where the model of energy spectrum $E_{u} \left(k\right)$ on large scale has been provided by \citep[see][Eq. (129)]{Xu:2022-The-statistical-theory-of-2nd}. In fact, constants $I_{m} $ are dependent on the scale factor \textit{a} since energy and momentum increase with time in self-gravitation collisionless flow (see Figs. \ref{fig:1a} and \ref{fig:3}), while they are time-independent in forced stationary turbulence. Nonetheless, constants $I_{m}$ reflect important dynamic properties on the large scale of SG-CFD. The physical significance of integral constant $I_{2} $ can be related to the linear momentum of entire system. 

For homogeneous velocity field with translational symmetry, the integral $\int _{V}^{}\left\langle \boldsymbol{\mathrm{u}}\cdot \boldsymbol{\mathrm{u}}^{'} \right\rangle d\boldsymbol{\mathrm{x}}^{3}  $ is independent of $\boldsymbol{\mathrm{x}}^{'} $ such that
\begin{equation} 
\label{ZEqnNum507946}
\begin{split}
I_{2}&=\int \left\langle \boldsymbol{\mathrm{u}}\cdot \boldsymbol{\mathrm{u}}^{'} \right\rangle d\boldsymbol{\mathrm{r}}^{3}  ={\mathop{\lim }\limits_{V\to \infty }} \frac{1}{V} \int _{V}\int _{V}^{}\left\langle \boldsymbol{\mathrm{u}}\cdot \boldsymbol{\mathrm{u}}^{'} \right\rangle d\boldsymbol{\mathrm{x}}^{3}   d\boldsymbol{\mathrm{x}}^{'3}\\
&={\mathop{\lim }\limits_{V\to \infty }} \frac{1}{V} \left\langle \int _{V}\int _{V}^{}\boldsymbol{\mathrm{u}}\cdot \boldsymbol{\mathrm{u}}^{'} d\boldsymbol{\mathrm{x}}^{3}   d\boldsymbol{\mathrm{x}}^{'3} \right\rangle={\mathop{\lim }\limits_{V\to \infty }} \frac{1}{V} \left\langle \int _{V}^{}\boldsymbol{\mathrm{u}}d\boldsymbol{\mathrm{x}}^{3}  \int _{V}\boldsymbol{\mathrm{u}}^{'}  d\boldsymbol{\mathrm{x}}^{'3} \right\rangle\\
&={\mathop{\lim }\limits_{V\to \infty }} \frac{1}{V} \left\langle \left(\int _{V}^{}\boldsymbol{\mathrm{u}}d\boldsymbol{\mathrm{x}}^{3}  \right)^{2} \right\rangle={\mathop{\lim }\limits_{V\to \infty }} V\left\langle \left(\frac{1}{V} \int _{V}^{}\boldsymbol{\mathrm{u}}d\boldsymbol{\mathrm{x}}^{3}  \right)^{2} \right\rangle,
\end{split}
\end{equation} 
where $\left\langle \bullet \right\rangle $ represents the ensemble average and \textit{V} is the volume of an enclosed system. With $V\to \infty $, effect of surface can be negligible and velocity field can be treated as a homogeneous field. 

If the system starts with a zero linear momentum that is conserved with time, i.e. $\int _{V}\boldsymbol{\mathrm{u}}d\boldsymbol{\mathrm{x}}^{3}  =0$, the integral constants $I_{2} =0$ is always true. This leads to a $k^{4} $ velocity spectrum on large scale if $I_{4} \ne 0$ (Eq. \eqref{ZEqnNum222875}), which is in alignment with the general argument about the influence of short scale gravitational interactions on large scales. The density power spectrum on scales much larger than the mean inter-particle spacing approaches a power-law $P_{\delta } \left(k\right)\sim k^{n} $ or $E_{\delta } \left(k\right)\sim k^{n+2} $ \citep{Baugh:1994-A-Comparison-of-the-Evolution-} (or equivalently $E_{u} \left(k\right)\sim k^{n} $ with $\delta \sim \nabla \cdot \boldsymbol{\mathrm{u}}$ on large scale ), where \textit{n} = 4 is the minimal large-scale power expected for discrete stochastic system \citep{Peebles:1980-The-Large-Scale-Structure-of-t}. Models for density and velocity correlations on large scale can be found in our previous work \citep[see][Section 5.1]{Xu:2022-The-statistical-theory-of-2nd}. 

The physical significance of integral constant $I_{4} $ is a little more complicated that can be briefly outlined here. Let's define the specific virial quantity \textit{G}, specific angular momentum \textbf{\textit{H}}, momentum tensor \textbf{\textit{M}},\textbf{ }and inertial tensor \textbf{\textit{I}} for an enclosed system with a volume \textit{V}, where $\boldsymbol{\mathrm{u}}$ is the peculiar velocity field and $\boldsymbol{\mathrm{x}}$ is the comoving coordinates from center of mass,
\begin{equation} 
\label{ZEqnNum272578} 
\begin{split}
&G=\frac{1}{V} \int _{V}^{}\boldsymbol{\mathrm{x}}\cdot \boldsymbol{\mathrm{u}}d\boldsymbol{\mathrm{x}}^{3}, \quad  \boldsymbol{\mathrm{H}}=\frac{1}{V} \int _{V}\boldsymbol{\mathrm{x}}\times \boldsymbol{\mathrm{u}}d\boldsymbol{\mathrm{x}}^{3},\\ 
&\boldsymbol{\mathrm{M}}=\frac{1}{V} \int _{V}\boldsymbol{\mathrm{x}}\otimes \boldsymbol{\mathrm{u}}d\boldsymbol{\mathrm{x}}^{3}, \quad \boldsymbol{\mathrm{I}}=\frac{1}{V} \int _{V}\boldsymbol{\mathrm{x}}\otimes \boldsymbol{\mathrm{x}}d\boldsymbol{\mathrm{x}}^{3}.  
\end{split}
\end{equation} 
Here \textit{G} and \textbf{\textit{H} }are defined\textbf{ }based on a continuum representation of velocity and density fields that is equivalent to the same quantities defined based on particle representation in N-body system (Eq. \eqref{ZEqnNum512486}), i.e. $G=G_{p} $ and $\boldsymbol{\mathrm{H}}=\boldsymbol{\mathrm{H}}_{p} $. For nonzero angular momentum $\boldsymbol{\mathrm{H}}\ne 0$, there exist nonzero off-diagonal terms in momentum tensor $\boldsymbol{\mathrm{M}}$ that comes from $\boldsymbol{\mathrm{H}}$, while the virial quantity \textit{G} is simply the trace of \textbf{M}. 

The integral constant $I_{4} $ can be written as (similar to Eq. \eqref{ZEqnNum530177}),
\begin{equation} 
\label{ZEqnNum461601}
\begin{split}
I_{4}&=\int \left\langle \boldsymbol{\mathrm{u}}\cdot \boldsymbol{\mathrm{u}}^{'} \right\rangle r^{2} d\boldsymbol{\mathrm{r}}^{3}\\
&={\mathop{\lim }\limits_{V\to \infty }} \frac{1}{V} \int _{V}\int _{V}^{}\left\langle \boldsymbol{\mathrm{u}}\cdot \boldsymbol{\mathrm{u}}^{'} \left(\boldsymbol{\mathrm{x}}^{'} -\boldsymbol{\mathrm{x}}\right)^{2} \right\rangle d\boldsymbol{\mathrm{x}}^{3}   d\boldsymbol{\mathrm{x}}^{'3}\\
&={\mathop{\lim }\limits_{V\to \infty }} \frac{1}{V} \int _{V}\int _{V}^{}\left\langle \boldsymbol{\mathrm{u}}\cdot \boldsymbol{\mathrm{u}}^{'} \left(-2\boldsymbol{\mathrm{x}}\cdot \boldsymbol{\mathrm{x}}^{'} \right)\right\rangle d\boldsymbol{\mathrm{x}}^{3}   d\boldsymbol{\mathrm{x}}^{'3},
\end{split}
\end{equation} 
where we use the fact linear momentum $\int _{V}^{}\boldsymbol{\mathrm{u}}d\boldsymbol{\mathrm{x}}^{3}=\int _{V}^{}\boldsymbol{\mathrm{u}}^{'} d\boldsymbol{\mathrm{x}}^{'3}  =0$ that requires
\begin{equation} 
\label{eq:80}
\begin{split}
\int _{V}\int _{V}^{}\left\langle \boldsymbol{\mathrm{u}}\cdot \boldsymbol{\mathrm{u}}^{'} \boldsymbol{\mathrm{x}}^{'} {}^{2} \right\rangle d\boldsymbol{\mathrm{x}}^{3}   d\boldsymbol{\mathrm{x}}^{'3}&=\int _{V}\int _{V}^{}\left\langle \boldsymbol{\mathrm{u}}\cdot \boldsymbol{\mathrm{u}}^{'} \boldsymbol{\mathrm{x}}^{2} \right\rangle d\boldsymbol{\mathrm{x}}^{3}   d\boldsymbol{\mathrm{x}}^{'3}\\
&=\left\langle \int _{V}^{}\boldsymbol{\mathrm{ux}}^{2} d\boldsymbol{\mathrm{x}}^{3}  \cdot \int _{V}\boldsymbol{\mathrm{u}}^{'}  d\boldsymbol{\mathrm{x}}^{'3} \right\rangle =0.
\end{split}
\end{equation} 

Note that Eq. \eqref{ZEqnNum461601} is only valid for an isolate enclosed system and not for system with periodic boundary. Extra care is needed for system with periodic boundary. To further reveal the physical significance of $I_{4} $, let's write two identities. The first identity is
\begin{equation} 
\label{eq:81} 
\left(\boldsymbol{\mathrm{x}}\times \boldsymbol{\mathrm{u}}\right)\cdot \left(\boldsymbol{\mathrm{x}}^{'} \times \boldsymbol{\mathrm{u}}^{'} \right)=\left(\boldsymbol{\mathrm{x}}\cdot \boldsymbol{\mathrm{x}}^{'} \right)\left(\boldsymbol{\mathrm{u}}\cdot \boldsymbol{\mathrm{u}}^{'} \right)-\left(\boldsymbol{\mathrm{x}}\cdot \boldsymbol{\mathrm{u}}^{'} \right)\cdot \left(\boldsymbol{\mathrm{x}}^{'} \cdot \boldsymbol{\mathrm{u}}\right).      
\end{equation} 
This can be easily proved using the notation of Einstein summation,
\begin{equation} 
\label{ZEqnNum763035} 
\begin{split}
\left(\boldsymbol{\mathrm{x}}\times \boldsymbol{\mathrm{u}}\right)\cdot \left(\boldsymbol{\mathrm{x}}^{'} \times \boldsymbol{\mathrm{u}}^{'} \right)&=x_{i} u_{j} \varepsilon _{ijk} x_{p}^{'} u_{q}^{'} \varepsilon _{pqk}\\ 
&=x_{i} u_{j} x_{p}^{'} u_{q}^{'} \left(\delta _{ip} \delta _{jq} -\delta _{iq} \delta _{jp} \right)\\
&=x_{i}^{} u_{j}^{} x_{i}^{'} u_{j}^{'} -x_{i}^{} u_{j}^{} x_{j}^{'} u_{i}^{'} \\
&=\left(\boldsymbol{\mathrm{x}}\cdot \boldsymbol{\mathrm{x}}^{'} \right)\left(\boldsymbol{\mathrm{u}}\cdot \boldsymbol{\mathrm{u}}^{'} \right)-\left(\boldsymbol{\mathrm{x}}\cdot \boldsymbol{\mathrm{u}}^{'} \right)\cdot \left(\boldsymbol{\mathrm{x}}^{'} \cdot \boldsymbol{\mathrm{u}}\right). 
\end{split}
\end{equation} 
Using the chain rule of differentiation, the second identity is,
\begin{equation}
\begin{split}
&\left(\boldsymbol{\mathrm{x}}\cdot \boldsymbol{\mathrm{x}}^{'} \right)\left(\boldsymbol{\mathrm{u}}\cdot \boldsymbol{\mathrm{u}}^{'} \right)+\left(\boldsymbol{\mathrm{x}}\cdot \boldsymbol{\mathrm{u}}^{'} \right)\cdot \left(\boldsymbol{\mathrm{x}}^{'} \cdot \boldsymbol{\mathrm{u}}\right)=x_{j}^{'} u_{i}^{'} \left(\delta _{ik} x_{j} +\delta _{jk} x_{i} \right)u_{k}\\
&=x_{j}^{'} u_{i}^{'} \left(x_{i} x_{j} \right)_{,k} u_{k} =x_{j}^{'} u_{i}^{'} \left(x_{i} x_{j} u_{k} \right)_{,k} -x_{j}^{'} u_{i}^{'} x_{i} x_{j} u_{k,k}. 
\end{split}
\label{ZEqnNum448940}
\end{equation}
\noindent Adding the two identities in Eqs. \eqref{ZEqnNum763035} and \eqref{ZEqnNum448940} leads to
\begin{equation} 
\label{ZEqnNum397887} 
\begin{split}
\left(\boldsymbol{\mathrm{x}}\times \boldsymbol{\mathrm{u}}\right)\cdot \left(\boldsymbol{\mathrm{x}}^{'} \times \boldsymbol{\mathrm{u}}^{'} \right)&=2\left(\boldsymbol{\mathrm{x}}\cdot \boldsymbol{\mathrm{x}}^{'} \right)\left(\boldsymbol{\mathrm{u}}\cdot \boldsymbol{\mathrm{u}}^{'} \right)\\
&-x_{j}^{'} u_{i}^{'} \left(x_{i} x_{j} u_{k} \right)_{,k} +x_{j}^{'} u_{i}^{'} x_{i} x_{j} u_{k,k}.
\end{split}
\end{equation} 

Integrating both sides of Eq. \eqref{ZEqnNum397887} with ${\int _{V}\int _{V}\left(\cdot \right)d\boldsymbol{\mathrm{x}}^{3} d\boldsymbol{\mathrm{x}}^{'3}/V^{2}}$ and taking average,
\begin{equation} 
\label{ZEqnNum134430} 
\begin{split}
\left|\boldsymbol{\mathrm{H}}\right|^{2}&=\boldsymbol{\mathrm{H}}\cdot \boldsymbol{\mathrm{H}}=2T-\frac{1}{V} \int _{V}x_{j}^{'} u_{i}^{'} d\boldsymbol{\mathrm{x}}^{'3} \cdot \frac{1}{V} \int _{V}\left(x_{i} x_{j} \right)_{,k} u_{k} d\boldsymbol{\mathrm{x}}^{3}\\  
&=\underbrace{2T}_{1}-\underbrace{\frac{1}{V} \int _{V}x_{j}^{'} u_{i}^{'} d\boldsymbol{\mathrm{x}}^{'3}  }_{3}\underbrace{\frac{1}{V} \int _{V}\left(x_{i} x_{j} u_{k} \right)_{,k} d\boldsymbol{\mathrm{x}}^{3}  }_{2}\\
&+\underbrace{\frac{1}{V} \int _{V}x_{j}^{'} u_{i}^{'} d\boldsymbol{\mathrm{x}}^{'3}  }_{3}\underbrace{\frac{1}{V} \int _{V}x_{i} x_{j} u_{k,k} d\boldsymbol{\mathrm{x}}^{3}  }_{4}\\ 
&=2T-\boldsymbol{\mathrm{M}}:\frac{1}{V} \int _{V}\left(x_{i} x_{j} u_{k} \right)_{,k} d\boldsymbol{\mathrm{x}}^{3}  +\left(\boldsymbol{\mathrm{M}}:\boldsymbol{\mathrm{I}}\right)\bar{u}_{k,k} ,
\end{split}
\end{equation} 
where term 1 is the contraction of the momentum tensor $\boldsymbol{\mathrm{M}}$,
\begin{equation}
\label{ZEqnNum611131} 
T=\frac{1}{V^{2} } \int _{V}\int _{V}^{}\left(\boldsymbol{\mathrm{x}}\cdot \boldsymbol{\mathrm{x}}^{'} \right)\left(\boldsymbol{\mathrm{u}}\cdot \boldsymbol{\mathrm{u}}^{'} \right)d\boldsymbol{\mathrm{x}}^{3}   d\boldsymbol{\mathrm{x}}^{'3} =\boldsymbol{\mathrm{M}}:\boldsymbol{\mathrm{M}}.         
\end{equation} 
The mean divergence $\bar{u}_{k,k} $ is defined as,
\begin{equation} 
\label{ZEqnNum509873} 
\bar{u}_{k,k} =\frac{1}{3} trace\left(\frac{\int _{V}x_{i} x_{j} u_{k,k} d\boldsymbol{\mathrm{x}}^{3}  }{\int _{V}x_{i} x_{j} d\boldsymbol{\mathrm{x}}^{3}  } \right).         
\end{equation} 
The second term in Eq. \eqref{ZEqnNum134430} is the integral of a divergence term. Using the divergence theorem, this term can be transformed into the integral of a flux on the surface enclosing the volume \textit{V}. The third term is the momentum tensor, and the fourth term is the inertial tensor weighted by velocity divergence. 

Equation \eqref{ZEqnNum134430} is formulated in a general setting that it can be applied to both isolate closed volume (halos) and periodic N-body system. 
From Eq. \eqref{ZEqnNum461601} , the physical meaning of integral constant $I_{4}$ for an isolate enclosed system can be related to scalar quantity $T$ as 
\begin{equation} 
\label{ZEqnNum958752} 
I_{4} =-2{\mathop{\lim }\limits_{V\to \infty }} \left(\left\langle T\right\rangle V\right).          
\end{equation} 
For incompressible flow with $u_{k,k} =0$ everywhere (term 4 in Eq. \eqref{ZEqnNum134430} vanishes), Eq. \eqref{ZEqnNum134430} reduces to the usual Landau-Loitsyansky equation \citep{Landau:1959-Fluid-Mechanics} for a closed volume with no surface flux contribution (term 2 in Eq. \eqref{ZEqnNum134430} should vanish), i.e.
\begin{equation}
\left\langle \left|\boldsymbol{\mathrm{H}}\right|^{2} \right\rangle =2\left\langle T\right\rangle \quad \textrm{and} \quad I_{4} =-{\mathop{\lim }\limits_{V\to \infty }} \left(\left\langle \left|\boldsymbol{\mathrm{H}}\right|^{2} \right\rangle V\right)     \label{ZEqnNum321115}
\end{equation}

\noindent from Eq. \eqref{ZEqnNum958752}, where the integral constant $I_{4}$ is related to the variance (fluctuation) of specific angular momentum $\boldsymbol{\mathrm{H}}$ of entire system. For incompressible turbulence, the conservation of angular momentum leads to a constant Loitsyansky integral $I_4$ such that the relation $u^{2} l^{5} =const$ can be applied to Eq. \eqref{ZEqnNum277600} for dynamics on large scale. 

It is entirely different in self-gravitating collisionless dark matter flow (SG-CFD). The irrotational nature and nonzero velocity divergence on large scale must be considered, which is different from incompressible flow \citep{Xu:2022-The-statistical-theory-of-2nd,Xu:2022-The-statistical-theory-of-3rd}. First, applying the divergence theorem leads to $\int _{V}\left(x_{i} x_{j} u_{k} \right)_{,k}  dV=0$, i.e. there is no net flow in or out of any enclosed system on large scale. Second, a relatively uniform divergence field can be assumed as $u_{k,k} \left(\boldsymbol{\mathrm{x}}\right)=\bar{u}_{k,k} $ because the density field $\delta \propto \nabla \cdot \boldsymbol{\mathrm{u}}$ is relatively uniform on the large scale. Now Eq. \eqref{ZEqnNum134430} can be reduced to  
\begin{equation} 
\label{ZEqnNum119840} 
\begin{split}
\left|\boldsymbol{\mathrm{H}}\right|^{2}&=2T+\left(\frac{1}{V} \int _{V}x_{j}^{'} u_{i}^{'} d\boldsymbol{\mathrm{x}}^{'3} \frac{1}{V} \int _{V}x_{i} x_{j} d\boldsymbol{\mathrm{x}}^{3}   \right)\bar{u}_{k,k}\\ &=2T+\left(\boldsymbol{\mathrm{M}}:\boldsymbol{\mathrm{I}}\right)\bar{u}_{k,k}.
\end{split}
\end{equation} 

On the large scale, the off-diagonal terms in momentum tensor \textbf{\textit{M}} should be negligible (the angular momentum is negligible on the large scale) such that (Eqs. \eqref{ZEqnNum272578} and \eqref{ZEqnNum611131}) we can write
\begin{equation} 
\label{ZEqnNum684864} 
\alpha _{T} T\approx G^{2} \gg \left|\boldsymbol{\mathrm{H}}\right|^{2},   
\end{equation} 
where the coefficient $\alpha _{T} $ should be related to the type of gravitational collapse on large scale. It can be easily confirmed that $\alpha _{T} =3$ for structure collapsing into a point ($\boldsymbol{\mathrm{u}}=-\boldsymbol{\mathrm{x}}$ everywhere), $\alpha _{T} =2$ for structure collapsing into a filament ($\boldsymbol{\mathrm{u}}=-{\mathrm{[x_1,x_2,0]}}$), and $\alpha _{T} =1$ for structure collapsing into a plane with $\boldsymbol{\mathrm{u}}=-{\mathrm{[x_1,0,0]}}$. 

Finally, different from hydrodynamic turbulence (Eq. \eqref{ZEqnNum321115}), for dark matter flow in an enclosed system, integral constant $I_{4} $ on large scale is related to the virial quantity as (from Eqs. \eqref{ZEqnNum958752} and \eqref{ZEqnNum119840}),
\begin{equation} 
\label{ZEqnNum121951} 
\begin{split}
I_{4}=-2{\mathop{\lim }\limits_{V\to \infty }} \left(\left\langle T\right\rangle V\right)&=-\frac{2}{\alpha _{T} } {\mathop{\lim }\limits_{V\to \infty }} \left(\left\langle G^{2} \right\rangle V\right)\\
&={\mathop{\lim }\limits_{V\to \infty }} \left(\left\langle \left(\boldsymbol{\mathrm{M}}:\boldsymbol{\mathrm{I}}\right)\bar{u}_{k,k} \right\rangle V\right).
\end{split}
\end{equation} 

The integral constants of motion ($I_4$ in Eq. \eqref{ZEqnNum121951}) can be applied to develop the models for large scale radial and angular momentum evolution. For a control volume (N-body simulation box) with a typical length scale (or size) \textit{L} and uniformly distributed mass, we expect the inertia matrix and virial quantity to be
\begin{equation}
\begin{split}
&\boldsymbol{\mathrm{I}}=\frac{1}{V} \int _{V}x_{i} x_{j} d\boldsymbol{\mathrm{x}}^{3}  =\frac{1}{3} \left(\frac{L}{2} \right)^{2} \delta _{ij}\\
&\textrm{and}\\
&G=\frac{1}{V} \int _{V}x_{j}^{} u_{i}^{} d\boldsymbol{\mathrm{x}}^{3} \delta _{ij} = \boldsymbol{\mathrm{M}}:\boldsymbol{\mathrm{\delta }},   \label{ZEqnNum424870}
\end{split}
\end{equation}

\noindent where $\delta _{ij} $ is a Kronecker delta. Substituting Eq. \eqref{ZEqnNum424870} into Eq. \eqref{ZEqnNum121951} for integral constant $I_4$ leads to the virial quantity \textit{G} that is dependent on the mean divergence defined in Eq. \eqref{ZEqnNum509873},
\begin{equation} 
\label{ZEqnNum342500} 
G=\frac{\alpha_{T}}{24} L^{2} \bar{u}_{k,k} .           
\end{equation} 
The average velocity divergence $\bar{u}_{k,k} $ can be related to the fluctuation of overdensity $\delta$ and the density fluctuation $\sigma _{\delta }^{2} $ on the scale of ${L/2}$ \citep[see][Eq. (119)]{Xu:2022-The-statistical-theory-of-2nd},
\begin{equation} 
\label{ZEqnNum625568} 
\left\langle \bar{u}_{k,k}^{2} \right\rangle =\left[aHf\left(\Omega _{m} \right)\right]^{2} \left\langle \delta ^{2} \right\rangle =\left[aHf\left(\Omega _{m} \right)\right]^{2} \sigma _{\delta }^{2} \left({L/2} \right),      
\end{equation} 
\begin{equation} 
\label{eq:96} 
\delta \approx \eta =-\frac{\nabla \cdot \boldsymbol{\mathrm{u}}}{aHf\left(\Omega _{m} \right)} \quad \textrm{and} \quad \left\langle \delta ^{2} \right\rangle = \sigma _{\delta }^{2}(L/2),          
\end{equation} 
where $f(\Omega_m)=1$ for matter dominant model with matter content $\Omega_m=1$.

The variance of density fluctuation (dispersion function) on a given scale \textit{r} is modelled previously \citep[see][Eq. (35)]{Xu:2022-Two-thirds-law-for-pairwise-ve}). On large scale, $\sigma _{\delta }^{2} $ is approximately modelled as,
\begin{equation}
\label{ZEqnNum784399} 
\sigma _{\delta }^{2} \left(r\right)\approx \frac{9}{2} \frac{a_{0} u^{2} r_{2}^{2} }{\left(aHf\left(\Omega _{m} \right)\right)^{2} r^{4} } ,         
\end{equation} 
where $u^{2} $ is one-dimensional velocity dispersion ($u_{0}^{2} $ is the velocity dispersion at present epoch). The parameter $a_{0} $ satisfies $a_{0} \left({u/u_{0} } \right)^{2} =0.45a$ \citep[see][Fig. 20]{Xu:2022-The-statistical-theory-of-2nd} and $r_{2} \approx 23{Mpc/h} $ is a constant comoving length sale that might be related to the size of sound horizon. Finally, the specific virial quantity reads,
\begin{equation}
{\mathop{\lim }\limits_{V\to \infty }} \left\langle G^{2} \right\rangle ={\mathop{\lim }\limits_{V\to \infty }} \left\langle G\right\rangle ^{2} =\frac{\alpha _{T}^{2} }{8} a_{0} u^{2} r_{2}^{2}, \quad \left\langle T\right\rangle =\frac{\alpha _{T} }{8} a_{0} u^{2} r_{2}^{2}. 
\label{ZEqnNum374769}
\end{equation}
Here the standard deviation of $G$ should vanish with $V\to \infty$ and Eqs. \eqref{ZEqnNum342500}, \eqref{ZEqnNum625568}, and \eqref{ZEqnNum784399} are used.

The variation of angular momentum on large scale for N-body system can be similarly modelled, 
\begin{equation}
\label{ZEqnNum748159} 
\left\langle H^{2} \right\rangle =0.002a_{0} u^{2} r_{2}^{2} a^{2} .         
\end{equation} 

\begin{figure}
\includegraphics*[width=\columnwidth]{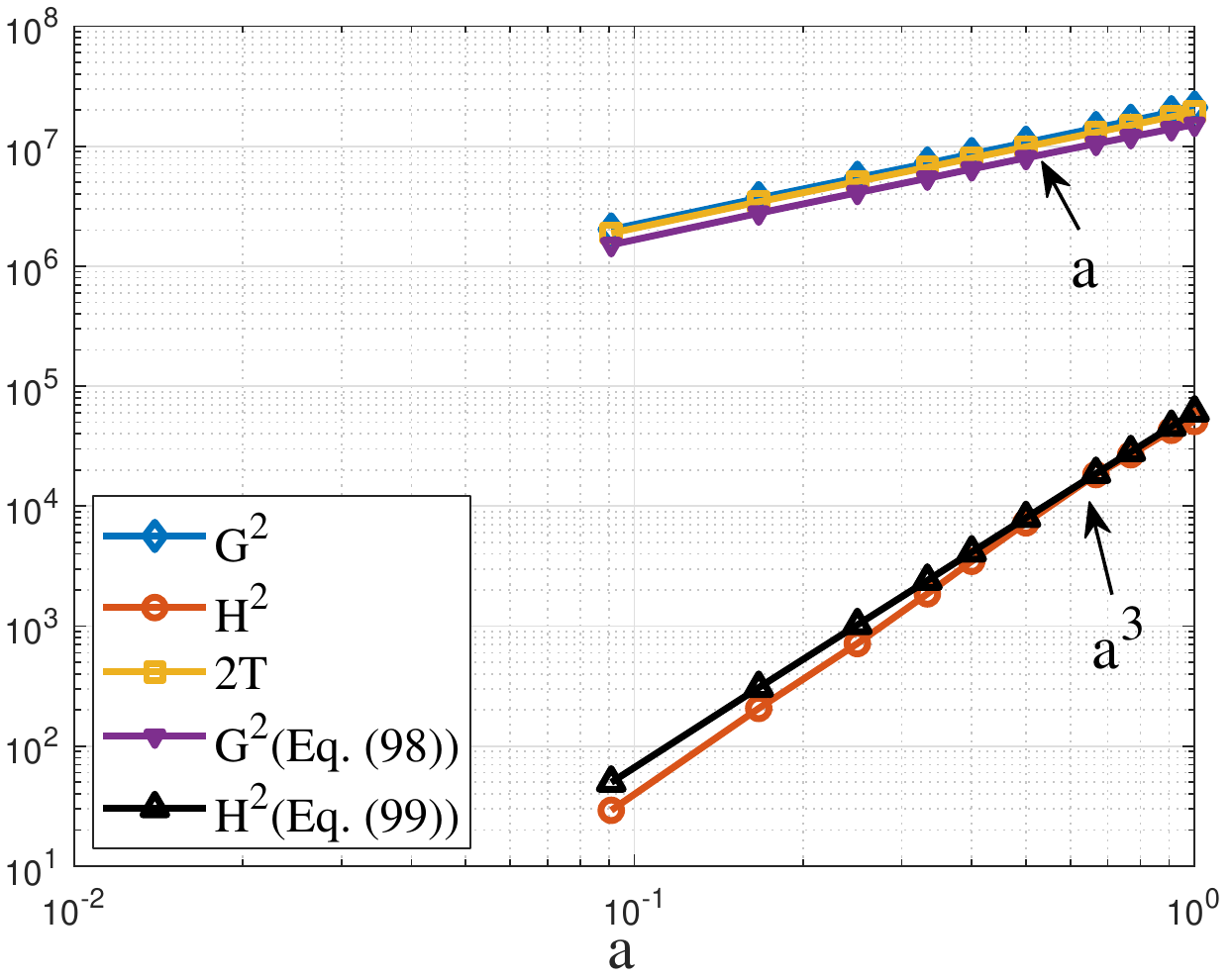}
\caption{The time variation of specific virial quantity $G^{2} $, angular momentum $\left|\boldsymbol{\mathrm{H}}\right|^{2} $, and  $T$ in Eq. \eqref{ZEqnNum134430} (unit: $(km/s\cdot Mpc/h)^2$) with scale factor \textit{a} from N-body simulation\textit{. }As expected, $G^{2} \propto a$ and $\alpha _{T} T\approx G^{2} $ with $\alpha _{T} \approx 2$ in Eq. \eqref{ZEqnNum684864}. The specific angular momentum $\left|\boldsymbol{\mathrm{H}}\right|\propto a^{{3/2} } \propto t\ll G$.}
\label{fig:7}
\end{figure}

Figure \ref{fig:7} plots the time variation of specific comoving virial quantity (radial momentum) $G^{2}$, angular momentum $\left|\boldsymbol{\mathrm{H}}\right|^{2} $, and  $T$ from N-body simulation in Section \ref{sec:2}, where $2T\approx G^{2}$ (Eq. \eqref{ZEqnNum684864}). In that case, $\alpha _{T} \approx 2$ (collapse into filament is the dominant mode in large scale N-body simulation), $G^{2} \propto a$ and $\left|\boldsymbol{\mathrm{H}}\right|^{2} \propto a^{3}$. The time variation of angular momentum $\left|\boldsymbol{\mathrm{H}}\right|^{2} $ from Eq. \eqref{ZEqnNum748159} is also presented in the same plot. Finally, all quantities on large scale can be related to dispersion $u^{2}$ and comoving length scale $r_{2}$ (Eqs. \eqref{ZEqnNum374769} and \eqref{ZEqnNum748159}). 

\subsection{Integral constants of motion on halo scale}
\label{sec:5.2}
On small (halo) scale, velocity field is of constant divergence \citep{Xu:2022-The-statistical-theory-of-2nd,Xu:2022-The-statistical-theory-of-3rd}. Equation \eqref{ZEqnNum134430} still can be applied to halos with an isothermal density profile, root mean square radius $r_{g}^{} $, specific angular momentum $\boldsymbol{\mathrm{H}}$\textbf{,} and virial quantity \textit{G}. The only differences from the derivation on large scale in previous section \ref{sec:5.1} are the non-uniform matter density on halo scale, constant velocity divergence, and non-vanishing surface term (term 2 in Eq. \eqref{ZEqnNum134430}). 

For halos with a coordinate $x_{3} $ aligned with the axis of spin, the momentum tensor reads \citep[see][Eq. (107)]{Xu:2022-The-mean-flow--velocity-disper}
\begin{equation} 
\label{eq:100} 
\boldsymbol{\mathrm{M}}=\frac{1}{m_{h} } \int _{V}\boldsymbol{\mathrm{x}}\otimes \boldsymbol{\mathrm{u}} \rho _{h} d\boldsymbol{\mathrm{x}}^{3} =\left[\begin{array}{ccc} {{G/3} } & {-{\left|\boldsymbol{\mathrm{H}}\right|/2} } & {0} \\ {{\left|\boldsymbol{\mathrm{H}}\right|/2} } & {{G/3} } & {0} \\ {0} & {0} & {{G/3} } \end{array}\right],       
\end{equation} 
such that
\begin{equation}
\begin{split}
&T=\boldsymbol{\mathrm{M}}:\boldsymbol{\mathrm{M}}=\left(\frac{1}{3} G^{2} +\frac{1}{2} \left|\boldsymbol{\mathrm{H}}\right|^{2} \right) \\
&\textrm{and}\\
&\alpha _{T} =\frac{G^{2} }{T} =\frac{6}{2+3{\left|\boldsymbol{\mathrm{H}}\right|^{2} /G^{2} } }.  
\end{split}
\label{eq:101}
\end{equation}
The diagonal terms in tensor $\boldsymbol{\mathrm{M}}$ are related to halo virial quantity \textit{G} while the off-diagonal terms are related to halo angular momentum $\left|\boldsymbol{\mathrm{H}}\right|$. Here $\alpha_T=3$ for non-rotating halos with radial flow only ($\boldsymbol{\mathrm{H}}=0$). The inertial tensor $\boldsymbol{\mathrm{I}}$ of a spherical halo is
\begin{equation} 
\label{eq:102} 
\boldsymbol{\mathrm{I}}=\frac{1}{m_{h} } \int _{V}x_{i} x_{j} \rho _{h} d\boldsymbol{\mathrm{x}}^{3}  =\frac{1}{3} r_{g}^{2} \delta _{ij} ,          
\end{equation} 
where $r_{g}^{} $ is the root mean square radius in Eq. \eqref{ZEqnNum753014}. For halos with a peculiar radial flow velocity $u_{r} =-Hr$ (\textit{H} is Hubble parameter), the velocity divergence is a constant with $u_{k,k} =-3H$. Similarly, the corresponding terms in Eq. \eqref{ZEqnNum134430} are 
\begin{equation} 
\label{eq:103} 
\begin{split}
\frac{1}{m_{h} } \int _{V}\left(x_{i} x_{j} \right)_{,k}  u_{k} \rho _{h} d\boldsymbol{\mathrm{x}}^{3}&=\frac{1}{m_{h} } \int _{V}\left(x_{i} u_{j} +x_{j} u_{i} \right)\rho _{h} d\boldsymbol{\mathrm{x}}^{3}\\  &=\boldsymbol{\mathrm{M}}+\boldsymbol{\mathrm{M}}^{T} =\frac{2}{3} G\delta _{ij},
\end{split}
\end{equation} 
and
\begin{equation} 
\label{eq:104} 
\frac{1}{m_{h} } \int _{V}x_{i} x_{j} u_{k,k} \rho _{h} d\boldsymbol{\mathrm{x}}^{3}  =-Hr_{g}^{2} \delta _{ij} .         
\end{equation} 
Therefore, we should have
\begin{equation} 
\label{ZEqnNum814432} 
\begin{split}
\frac{1}{m_{h} } \int _{V}\nabla \cdot \left(\boldsymbol{\mathrm{x}}\otimes \boldsymbol{\mathrm{x}}\otimes \boldsymbol{\mathrm{u}}\right)\rho _{h} d\boldsymbol{\mathrm{x}}^{3}  &=\frac{1}{m_{h} } \int _{V}\left(x_{i} x_{j} u_{k} \right)_{,k} \rho _{h} d\boldsymbol{\mathrm{x}}^{3} \\
&=\left(\frac{2}{3} G-Hr_{g}^{2} \right)\delta _{ij}.
\end{split}
\end{equation} 

With Eqs. \eqref{eq:101}, \eqref{eq:102}, and \eqref{ZEqnNum814432}, Eq. \eqref{ZEqnNum134430} is automatically satisfied for isothermal halos. At the same time with peculiar radial flow $u_{r} =-Hr$, we have
\begin{equation}
\label{ZEqnNum530165} 
\frac{1}{m_{h} } \int _{V}\nabla \cdot \left(\boldsymbol{\mathrm{x}}\otimes \boldsymbol{\mathrm{x}}\otimes \boldsymbol{\mathrm{u}}\right)\rho _{h} d\boldsymbol{\mathrm{x}}^{3}  =-5H\boldsymbol{\mathrm{I}}=-\frac{5}{3} Hr_{g}^{2} \delta _{ij} .      
\end{equation} 
Comparing Eqs. \eqref{ZEqnNum814432} and \eqref{ZEqnNum530165} leads to an expression for isothermal halo \citep[also see][Eq. (103)]{Xu:2022-The-mean-flow--velocity-disper},
\begin{equation} 
\label{ZEqnNum738790} 
G=-Hr_{g}^{2} .           
\end{equation} 
From Eq. \eqref{ZEqnNum958752}, finally, the integral constant of motion $I_{4}$ on halo scale is related to both specific virial quantity and angular momentum
\begin{equation} 
\label{ZEqnNum707811} 
I_{4} =-V\left(\frac{2}{3} G^{2} +\left|\boldsymbol{\mathrm{H}}\right|^{2} \right),         
\end{equation} 
while $I_{4}$ is only related to the virial quantity on large scale (Eq. \eqref{ZEqnNum121951}).

\section{Conclusions}
\label{sec:6}
The energy and momentum evolution are studied on both large and halo scales for self-gravitating collisionless dark matter flow (SG-CFD) in expanding background. The gravitational interaction is assumed to have an arbitrary potential exponent \textit{n}. Instead of working with the original comoving system, the temporal evolution is formulated in a transformed system with a static background and fixed damping for convenience (Eqs. \eqref{ZEqnNum333845}, \eqref{ZEqnNum672411} and \eqref{ZEqnNum475585}). The equivalence between transformed and original comoving systems is established. 

Regardless of the potential exponent \textit{n}, the energy equation in transformed system has a simple form that is identical with the energy evolution of a damped harmonic oscillator (Eq. \eqref{ZEqnNum333845}). To solve the energy equation, additional insights are required to complete the energy equation. This can be supplemented by an elementary problem of gravitational collapse, i.e. a two-body collapse (TBCM) model. The TBCM model predicts an exponential evolution of system energy in transformed system, or equivalently a power-law evolution in a comoving system. Therefore, the energy equation of SG-CFD admits a power-law evolution of kinetic and potential energy as exact steady-state solutions that are consistent with N-body simulation (Eqs. \eqref{ZEqnNum840626}-\eqref{ZEqnNum680744} and Fig. \ref{fig:1a}). 

This suggests an effective potential exponent $n_{e}$ (Eqs. \eqref{ZEqnNum539248} and \eqref{ZEqnNum751621}) to incorporate the dynamic effects of mass cascade, where $n_{e} $ can be different from the usual gravitational potential of \textit{n }=-1. Simulation suggests $n_{e} =-1.38$ (Fig. \ref{fig:1b}), while the energy evolution for typical halos (grow with a constant waiting time) suggests $n_{e} =-{10/7} $. With  $n_{e} =-{10/7} $ (also predicted by TBCM model \citep[see][Eq. (83)]{Xu:2021-A-non-radial-two-body-collapse}, the kinetic and potential energy of entire N-body system increase linearly with time. A constant energy production rate $\varepsilon _{u} $ can be introduced (Eq. \eqref{ZEqnNum840626}) that is consistent with the picture of inverse energy cascade \citep{Xu:2021-Inverse-and-direct-cascade-of-}, where kinetic energy is injected on the smallest scale at a constant rate and transferred to larger scales. Unlike the freely decaying turbulence, kinetic energy in SG-CFD constantly grows due to continuous virilization in halos. 

For any halo with a given mass $m_{h} $ identified in simulation, the halo virial kinetic energy $\sigma _{v}^{2} $ and root mean square radius $r_{g} $ can be unambiguously determined (Eq. \eqref{ZEqnNum753014}). Two constants $\alpha _{s}^{*} $ and $\beta _{s}^{*} $ are introduced for each halo to relate halo size $r_{g} $,  kinetic energy $\sigma _{v}^{2} $, and mass $m_{h} $ (Eq. \eqref{ZEqnNum569238}). The mean values of both constants for all halos with same mass (i.e. in the same halo group) are relatively independent of redshift and halo mass and estimated to be $\beta _{s}^{*} \approx 0.09$ and $\alpha _{s}^{*} \approx 0.29$ (Eq. \eqref{ZEqnNum459478}). However, small halos tend to have a greater standard deviation in both constants. Large halos tend to be synchronized and generated at the same time with smaller dispersion in halo properties (Fig. \ref{fig:2}). Just like the exponent $n_{e} $ for entire system, an effective potential exponent $n_{s}^{*} $ (or virial ratio $\gamma _{v} =-n_{s}^{*} $)  can be introduced for each halo with $n_{s}^{*} \approx -1.3$ for large halos (Fig. \ref{fig:5}). 

The momentum evolution of N-body system is also formulated (Eqs. \eqref{ZEqnNum257929} and \eqref{ZEqnNum475585}). N-body simulation suggests a scaling of comoving virial quantity (radial momentum) $G_{p} \propto a^{{1/2} }$ and angular momentum $H_{p} \propto a^{{3/2} } $ (Fig. \ref{fig:3}). The entire system can be divided into a halo sub-system including all particles in all halos and an out-of-halo sub-system including all particles not in any halos. The momentum of entire halo sub-system can be decomposed into contributions from motion of particles in halos and motion of halos (Eq. \eqref{ZEqnNum727626}). The averaged specific virial quantity $\left\langle G_{h} \right\rangle $ and angular momentum $\left\langle \left|\boldsymbol{\mathrm{H}}_{h} \right|\right\rangle$ (in physical coordinate) for entire halo sub-system has a scaling of $\propto a^{{3/2}}$ (Fig. \ref{fig:3}). 

Momentum in each halo can be also conveniently modeled in terms of  halo kinetic energy $\sigma _{v}^{2} $ and root mean square radius $r_{g} $ by introducing two mass-dependent coefficients $\eta _{s}^{*} \left(m_{h} \right)$ and $\tau _{s}^{*} \left(m_{h} \right)$ (Eq. \eqref{ZEqnNum336536} and Fig. \ref{fig:6}). The limiting values of $\eta _{s}^{*} $ and $\tau _{s}^{*} $ for small and large halos are also identified (Eqs. \eqref{ZEqnNum292188}-\eqref{ZEqnNum112040}). The halo spin parameter $\lambda _{p} $ can be expressed in terms of parameters $\alpha _{s}^{*} $, $\eta _{s}^{*} $, and potential exponent $n_{s}^{*}$ (Eq. \eqref{ZEqnNum879931}), where $\lambda _{p} $ decreases with halo mass. For small halos with very slow mass accretion, $\lambda _{p} $ increases with time $\propto a^{{1/2} } $, while for large halos in their early stage with fast mass accretion, $\lambda _{p} \approx 0.031$ (Fig. \ref{fig:6} and Eq. \eqref{ZEqnNum879931}). All relevant parameters for halo momentum, energy and spin parameter are summarized in Table \ref{tab:3} with limiting values for small and large halos. 

Finally, a series of integral constants of motion can be introduced to describe the large-scale dynamics for SG-CFD. The \textit{m}th order constant $I_{m} $ is essentially the integral of \textit{(m-2)}th moments of velocity correlation function (Eq. \eqref{ZEqnNum530177}) and is related to the \textit{m}th derivative of energy spectrum at small \textit{k} or long wavelength limit (Eq. \eqref{ZEqnNum222875}). The comoving virial quantity \textit{G} and angular momentum \textbf{H} of entire N-body system (Eq. \eqref{ZEqnNum272578}) are intimately related to these integral constants. Here $I_{2} $ is related to the fluctuation of linear momentum (Eq. \eqref{ZEqnNum507946}) and should vanish due to the conservation of linear momentum. The constant $I_{4}$ is related to the fluctuation of virial quantity (radial momentum) of SG-CFD (Eq. \eqref{ZEqnNum121951}), where angular momentum is relatively small on large scale, i.e. $\left|\boldsymbol{\mathrm{H}}\right|\ll G$. By contrast, $I_{4} $ is related to the angular momentum that is dominant in hydrodynamic turbulence. The evolution of virial quantity $G^{2} $ and angular momentum $\left|\boldsymbol{\mathrm{H}}\right|^{2} $ can be modelled on large scale and compared with simulation (Eqs. \eqref{ZEqnNum374769}, \eqref{ZEqnNum748159} and Fig. \ref{fig:7}). On halo scale, the constant $I_{4} $ is related both $G^{2} $ and $\left|\boldsymbol{\mathrm{H}}\right|^{2} $, both of which are comparable (Eq. \eqref{ZEqnNum707811}).


\section*{Data Availability}
Two datasets underlying this article, i.e. a halo-based and correlation-based statistics of dark matter flow, are available on Zenodo \citep{Xu:2022-Dark_matter-flow-dataset-part1,Xu:2022-Dark_matter-flow-dataset-part2}, along with the accompanying presentation slides "A comparative study of dark matter flow \& hydrodynamic turbulence and its applications" \citep{Xu:2022-Dark_matter-flow-and-hydrodynamic-turbulence-presentation}. All data files are also available on GitHub \citep{Xu:Dark_matter_flow_dataset_2022_all_files}.

\bibliographystyle{mnras}
\bibliography{Papers}


\label{lastpage}
\end{document}